\documentclass[12pt]{iopart}
\usepackage{iopams}  
\usepackage{url}
\usepackage{cleveref} %
\usepackage{graphicx} %
\usepackage{color}
\usepackage{xr}
\usepackage[utf8]{inputenc}
\usepackage{bm} %
\usepackage{pdfpages} %

\newcommand{\db}{\mathbf{d}}                                                                                         
\newcommand{\eb}{\mathbf{e}}

\newcommand{\gb}{\mathbf{g}}                                                                                         
\newcommand{\Hb}{\mathbf{H}}                                                                                         

\newcommand{\nub}{\bm{\nu}}
\newcommand{\ub}{\mathbf{u}}
\newcommand{\vb}{\mathbf{v}}
                                                                                         
\newcommand{\xb}{\mathbf{x}}        

\newcommand{\nablab}{\bm{\nabla}}  %

\crefname{equation}{equation}{equations}
\Crefname{equation}{Equation}{Equations}

\newcommand{\yb}{\mathbf{y}}                                                                                         
\newcommand{\Pb}{\mathbf{P}}

\newcommand{\sigmam}{\lambda_m}
\newcommand{\sigmat}{\lambda_n}
\newcommand{\sigmar}{\lambda_r}                                                                                      

\newcommand{\B}{B} %
\newcommand{\R}{\mathbb{R}} %
\newcommand{\nb}{n_B} %
\newcommand{\nd}{n_d} %
\newcommand{\nx}{n_x} %

\newcommand{\cmmnt}[1]{\ignorespaces}

\definecolor{dgreen}{rgb}{0.0,0.6,0.0}

\newcommand{\revised}[2]{#2}

\newcommand{\titlestringshort}{{Algorithms for Training Physics Models}}
\newcommand{\titlestringlong}{{Optimization and Supervised Machine Learning Methods for Fitting Numerical Physics Models without Derivatives}}

\DeclareGraphicsExtensions{.pdf}
\graphicspath{{./images/}}

\begin{document}

\title[\titlestringshort]{\titlestringlong}

\author{Raghu Bollapragada$^{1,2}$, Matt Menickelly$^1$, Witold~Nazarewicz$^{3}$, Jared O'Neal$^1$, Paul-Gerhard Reinhard$^4$, Stefan M. Wild$^1$}

\address{$^1$ Mathematics and Computer Science Division, Argonne National Laboratory, Lemont, Illinois 60439, USA}
\address{$^2$
Operations Research and Industrial Engineering Graduate Program, Department of Mechanical Engineering, University of Texas, Austin, Texas 78712, USA}
\address{$^3$
Department  of  Physics  and  Astronomy  and  FRIB  Laboratory, Michigan  State  University,  East  Lansing,  Michigan  48824,  USA}
\address{$^4$ Institut  f\"ur  Theoretische  Physik  II,  Universit\"at  Erlangen-N\"urnberg,  D-91058  Erlangen,  Germany}
\ead{wild@anl.gov}

\begin{abstract}
We address the calibration of a computationally expensive nuclear physics model for which derivative information with respect to the fit parameters is not readily available. Of particular interest is the performance of optimization-based training algorithms when dozens, rather than millions or more, of training data are available and when the expense of the model places limitations on the number of concurrent model evaluations that can be performed.

As a case study, we consider the Fayans energy density functional model, which has characteristics similar to many model fitting and calibration problems in nuclear physics. We analyze hyperparameter tuning considerations and variability associated with stochastic optimization algorithms and illustrate considerations for tuning in different computational settings.

\end{abstract}

\vspace{2pc} 


\section{Introduction}
\label{sec:intro}
A problem arising throughout both nuclear theory---from {\it ab-initio} nuclear
theory \cite{Ekstrom2019,Piarulli2016} to density functional theory (DFT)
\cite{Bender2003,schunck2019energy}---and supervised machine learning is the
fitting of a model to data. Formally, given a computer model $m$ evaluated at
inputs $\nub_1, \ldots, \nub_{\nd}$, one seeks a parameter vector $\xb \in \R^{\nx}$ so that
the outputs $m\left(\nub_1;\xb\right), \ldots, m\left(\nub_{\nd};\xb\right)$
agree with data $\db=[d_1, \ldots, d_{\nd}]$ within the assumed uncertainties. For example, the inputs $\nub$
might characterize a particular  configuration of the atomic nucleus (defined by   the number of its
proton and neutron constituents).  For such a case, the
data might include observables such as experimentally measured binding energies and  charge
radii. Or, generally, the inputs could correspond to 64-pixel by 64-pixel images, and the data could represent labels such as ``cat'' or ``banana.'' 

Although fitting can result in many different formulations of optimization problems, the most common form in 
the physical sciences 
follows a $\chi^2$-convention wherein independence of errors is assumed and one seeks to solve
\begin{equation}
\label{eq:function}
\min_{\xb \in \R^{\nx}} f(\xb), \qquad \mbox{where } f(\xb) = \sum_{i=1}^{\nd} \left(\frac{m\left(\nub_i;\xb\right) - d_i}{\sigma_i}\right)^2,
\end{equation}
with $\sigma_1, \ldots, \sigma_{\nd}>0$ often interpreted as experimental and/or model error bars \cite{DNR14}. 
More general types of such objective functions (also called ``loss functions'' or ``penalty functions'') include those that take into account correlations, such as
\begin{equation*}
\label{eq:correlated}
f(\xb) = \sum_{i=1}^{\nd} \sum_{j=1}^{\nd} w_{i,j} \left(m\left(\nub_i;\xb\right) - d_i\right) \left(m\left(\nub_j;\xb\right) - d_j\right).
\end{equation*}

In this paper we address the squared-loss formulation in \cref{eq:function},
which we generalize as the finite sum of squares of nonlinear functions of the
parameter vector $\xb$; that is,
\begin{equation}
\label{eq:genfun}
f(\xb) = \sum_{i=1}^{\nd} F_i(\xb)^2.
\end{equation}
Throughout the text, we refer to these general functions, $F_i$, as \emph{component functions}.
Since objective functions of the form in \cref{eq:genfun} are found throughout supervised learning, 
many optimization methods used for training machine learning models are applicable here.
In contrast to
standard fitting problems that arise in nuclear theory, however, the number of data,
$\nd$, used when training machine learning models tends to be massive. For
example, as of August 2020, the open images dataset 
\cite{openimages20}
contained nearly 60 million image labels.  
When fitting nuclear models, 
the value of $\nd$ is typically many orders of magnitude smaller; this is the case in the study conducted in this paper. 

A natural question is thus whether the algorithms used to train machine learning models can benefit the physicist who has a computer model and desires to solve fitting problems. 
Here we investigate the strengths and limitations of different optimization algorithms for minimizing \cref{eq:genfun} through a case study from nuclear theory. We focus on the ``derivative-free'' case where gradients 
$\nablab f(\xb), \nablab F_1(\xb),  \ldots, \nablab F_{\nd}(\xb)$ 
and higher-order derivatives 
are unavailable for use in the optimization. 
This is often the setting when the computer models are composed of iterative components \cite{more2014nd,WSS15} or when dependencies on legacy computer codes pose obstacles to algorithmic differentiation \cite{Berz1996}, which is a key enabling technology for deep learning \cite{2015arXiv150205767G}.

In \cref{sec:algorithms} we summarize the set of optimization algorithms tested. We focus on methods for local optimization (i.e., those that do not seek to asymptotically cover the entire parameter space) since we are interested in assessing performance within a budget of function evaluations.
Such a budget limits the applicability of global optimization algorithms. 
Our case study, involving the fitting of the Fayans energy density functional (EDF) to
data across the nuclear chart, is described in \cref{sec:fayans}. This problem was selected in part because it shares characteristics with many fitting problems.
For this problem, there are $\nd=198$ data, $\nx=13$ parameters to be optimized, and correlations among the errors $m\left(\nub_i;\xb\right) - d_i$ are evident.
Numerical results are presented in \cref{sec:numerical}, and we summarize both the
consistency and efficiency of the tested algorithms. Our performance measures
emphasize how the efficiency of optimization methods, as measured in function evaluations, can change depending on one's ability to evaluate components $F_i(\xb)$ concurrently. 

Although our focus is on optimization-based approaches for training, these could also be used in a larger framework of statistical calibration (e.g., as discussed in \cite{HigdonJPG15}).

\section{Derivative-free optimization/training algorithms}
\label{sec:algorithms}

We consider five algorithmic families of iterative methods for local, unconstrained derivative-free optimization. 
The first two algorithms are deterministic, and the latter three are randomized
(sometimes called ``stochastic''). For randomized algorithms, the 
sequence of points in parameter space at which the component functions will be evaluated is generated stochastically/nondeterministically by the method. 

The randomized algorithms considered in this study are designed to have the ability to vary the number of component functions evaluated in any one iteration. 
Throughout this paper, we refer to this number as the \emph{batch size}, denoted $\nb$.
In our experiments, as is typical in such batch-sampling-based algorithms, a
batch of size $\nb$ is generated by sampling uniformly $\nb$ many times without
replacement from the integers $\{1, \ldots, \nd\}$. Hence, the maximum batch size $\nb=\nd$ corresponds to evaluating all of the component functions.

We now briefly describe each of the algorithm types, along with their hyperparameters and our implementation. 
For additional details on these and other derivative-free optimization methods, we refer the reader to \cite{Conn2009a,LMW2019AN}.

\subsection{Deterministic algorithms}

In general, deterministic methods have the property that, given a starting point
and hyperparameter values, the sequence of points in parameter space generated by the method will be
the same every time the optimization is repeated. The deterministic methods considered here also assume that all of the
$\nd$ component functions in \cref{eq:genfun} are evaluated before the next
point in parameter space to be evaluated is determined. 
That is, we address a batch-synchronous, rather than asynchronous, environment; the latter may be appropriate when the component function evaluation time varies significantly and/or individual component functions depend on relatively few parameters \cite{Recht2011hogwild}.

\subsubsection{Direct search algorithm.}
\label{sec:nelder}

The Nelder-Mead simplex algorithm \cite{NelderMead} is a popular direct search algorithm for general derivative-free optimization \cite{PhysRevLett.67.1334,pres07}. The version tested here is from the MATLAB routine \texttt{fminsearch} based on \cite{JCL98}. 

The Nelder-Mead algorithm determines a new point for evaluation by performing operations on a simplex
defined by $\nx+1$ previously evaluated affinely independent points
\cite{Conn2009a,Wright2012,AudetHare2017}. The particular choice of operations is dictated 
by the $f$ values associated with each of the simplex's vertices. 
Since the algorithm bases its decision on the complete evaluation of $f$, accepted points $\xb_k$ monotonically decrease the objective: $f(\xb_0) > f(\xb_1) > \ldots$. 
Multiple complete evaluations of $f$ may be required before an acceptable point is found, and each of these evaluations corresponds to $\nd$ component function evaluations.

The sole hyperparameter in our Nelder-Mead implementation is the initial simplex size. This size can be interpreted as defining the size of the neighborhood within which Nelder-Mead begins its search. 

\subsubsection{Model-based trust-region algorithm.}
\label{sec:pounders}

POUNDERS \cite{SWCHAP14} is a deterministic method that exploits the structural form in \cref{eq:genfun} by constructing a local surrogate model of each component function $F_i$. POUNDERS was used in the optimization of the UNEDF family of energy density functionals \cite{UNEDF0,UNEDF1,UNEDF2} and chiral nucleon-nucleon interactions \cite{Ekstrom13,EksJPG15}.  

The surrogate models in POUNDERS are updated with each new function evaluation, and the algorithm assumes that all $\nd$ component functions are evaluated at each point. 
A new point to evaluate 
is obtained by locally minimizing an
aggregation of the component surrogate models.  Thus, unlike the Nelder-Mead
method, POUNDERS requires and exploits full knowledge of the individual 
component function values $F_1(\xb_k), \ldots, F_{\nd}(\xb_k)$. 
Similar to Nelder-Mead, 
since POUNDERS evaluates all component functions, accepted points monotonically decrease the objective, and multiple such evaluations of $\nd$ components may be required before decrease in the function value is found.

The primary hyperparameter in POUNDERS is the radius used to define the initial neighborhood within which the surrogate models are constructed and optimized over.

\subsection{Derivative-free stochastic approximation}
\label{sec:sgd}

In supervised learning tasks of machine learning---a class of optimization problems containing \cref{eq:genfun}---the
workhorse optimization method for obtaining (approximate) solutions to \cref{eq:genfun} has been 
stochastic gradient methods \cite{RobbinsMonro1951}. 
For an excellent contemporary survey of stochastic gradient methods, see \cite{BottCurtNoce16}. 
Stochastic gradient methods as applied to \cref{eq:genfun} resemble traditional gradient descent methods, 
the basic iteration of which takes the form
\begin{equation}\label{eq:gd_iteration}
 \xb_{k+1}= \xb_k-\alpha_k\nablab f(\xb_k).
\end{equation}
When $\nd$ is large, however, it may become computationally prohibitive to
evaluate, or even numerically estimate, the gradient of \cref{eq:genfun}:
\begin{equation}
\label{eq:gradsum}
\nablab f(\xb_k)=\sum_{i=1}^{\nd} \nablab F_i^2(\xb_k).
\end{equation}

Thus, stochastic gradient methods compute approximations to the gradient  \cref{eq:gradsum} 
by including in the sum only a
sampled batch of the component function indices $\{1,\dots,\nd\}$. 
In its simplest form, a single $i(k)\in\{1,\dots,\nd\}$ is chosen at random from a 
discrete (often uniform) distribution, and the basic iteration in \cref{eq:gd_iteration} is replaced with
\begin{equation}\label{eq:sgd_basic_iteration}
 \xb_{k+1}= \xb_k-\alpha_k\nd\nablab F_{i(k)}^2(\xb_k).
\end{equation}
Effectively, the cost of performing  \cref{eq:sgd_basic_iteration}
is a factor of $\nd$ times cheaper than the cost of performing \cref{eq:gd_iteration}.
This represents significant computational savings in performing a single iteration when $\nd$ is large,
at the expense of using an inaccurate gradient approximation.
More generally, one can consider sampling a batch of component function indices
$\B_k\subseteq\{1,\dots,\nd\}$ of size $\nb(k)\leq\nd$,
and replacing \cref{eq:gd_iteration} with 
\begin{equation}\label{eq:sgd_batch_iteration}
 \xb_{k+1} = \xb_k-\alpha_k\displaystyle\frac{\nd}{\nb(k)}\displaystyle\sum_{i\in B_k}\nablab F_{i}^2(\xb_k).
\end{equation}
The rationale behind the random sampling approach in \cref{eq:sgd_batch_iteration}
is that the expected value (with
respect to the stochastic sampling of component function indices) of 
$\displaystyle\frac{\nd}{\nb(k)}\displaystyle\sum_{i\in B_k}\nablab F_{i}^2(\xb_k)$
is exactly
\cref{eq:gradsum}.  
We note that when $\nb(k) < \nd$, the step from $\xb_k$ to $\xb_{k+1}$ will be based on incomplete information; however, since the sampled batches will be independent from one iteration to the next, these methods probabilistically find a zero of the full gradient \cref{eq:gradsum} when the step sizes decay fast enough. 

Dating almost as far back as the earliest stochastic gradient methods \cite{RobbinsMonro1951},
derivative-free variants of the iterations in \cref{eq:sgd_basic_iteration} and in \cref{eq:sgd_batch_iteration} have been proposed \cite{KieferWolfowitz}. 
All these methods 
perform an analogous iteration 
\begin{equation}
\label{eq:basic_update}
\xb_{k+1} = \xb_k - \alpha_k \db_k, \qquad k=0,1,2,\ldots,
\end{equation}
with $\db_k \in \R^{\nx}$ serving as an estimate of a gradient quantity. 
These algorithms differ in their selection of the step size $\alpha_k>0$ and, most distinctively, the choice of direction $\db_k$. 

\subsubsection{Kiefer-Wolfowitz method.}
Iterations of type \cref{eq:basic_update} are found in the Kiefer-Wolfowitz (KW) method \cite{KieferWolfowitz}.
The KW method computes \emph{finite differences of sampled functions} to
approximate directional derivatives in each of the $\nx$ coordinate directions. 
Although other variants exist  \cite{LEcuyerYin1998,Kleinman1999}, 
in this paper we use the most common sampling described in the following.

In the $k$th iteration, we uniformly sample a 
batch $B_k\subseteq\{1,\dots,\nd\}$ of size $|B_k|=\nb(k)$.  
Given a fixed finite-difference parameter $h$, 
forward differences are used to approximate the partial derivatives needed to
estimate the gradients of the $\nb(k)$ squared component functions associated
with the batch.  Specifically, we compute
\begin{equation}
\label{eq:kw-fs}
\db_k = \frac{\nd}{\nb(k)} \displaystyle  \sum_{i\in\B_k} 
\gb_i(\xb_k;h), 
\end{equation}
where
\begin{equation}
\label{eq:FD}
\gb_i(\xb_k;h) = \frac{1}{h}
\left[\begin{array}{c}
F_i\left(\xb_k+h\eb_1\right)^2-F_i\left(\xb_k\right)^2 \\
\vdots \\
F_i\left(\xb_k+h\eb_{\nx}\right)^2-F_i\left(\xb_k\right)^2 
\end{array}\right].
\end{equation}
In our experiments, we refer to the algorithm that uses \cref{eq:kw-fs} as $\db_k$ in \cref{eq:basic_update}
as ``KW.'' 

Observe that in KW, $\nb(\nx+1)$ component function evaluations are performed in a single iteration. 
As with any method using \cref{eq:basic_update}, both a sequence of step sizes $\{\alpha_k\}$ and a sequence
of batch sizes $\{\nb(k)\}$ must be selected. 
Additionally, the finite-difference parameter $h>0$ must be selected. 
For the sake of simplicity in presentation,
we have chosen to keep $h$ fixed, with the immediate consequence that 
$\db_k$ is a biased estimator of $\nablab f(\xb_k)$  for all $k$, even when $\nb(k)=\nd$. 

\subsubsection{Bandit method.}
\label{sec:bandit}
We now consider members of a class of derivative-free methods that have become increasingly attractive in supervised learning over the past decade,
the so-called (two-point) bandit methods \cite{Agarwal2010,Ghadimi2013,Duchi2015,Gasnikov2017,Shamir2017}. 
Similar to KW, bandit methods can employ a batch
$B_k\subseteq\{1,\dots,\nd\}$ of component function indices in the computation
of $\db_k$, which is computed based on finite differences and employed in
iterations of the type \cref{eq:basic_update}.
In each iteration of a bandit method, however, only \emph{one} directional
derivative is numerically approximated per element in the batch; in contrast, KW uses 
$\nx$ partial derivatives per element.
For the basic iteration of what we refer to as the ``Bandit'' method in the following,
the direction vector $\db_k$ becomes
\begin{equation}
\label{eq:bandit-fs}
\db_k = \frac{\nd}{\nb(k)} \sum_{i\in\B_k} 
\left(
\frac{F_i\left(\xb_k+h\ub_k\right)^2-F_i\left(\xb_k\right)^2}{h} 
\right)\ub_k,
\end{equation}
where $\ub_k$ is a \emph{randomized} direction.  In particular, 
we sample $\ub_k$ uniformly from the surface of an $\nx$-dimensional sphere centered at the origin and of unit radius.
Once again, the quantity $h$ in \cref{eq:bandit-fs} denotes a fixed finite-difference parameter. 
In the case where $\nb(k)=\nd$ for all $k$, the Bandit method is related to the
iteration used in the Gaussian smoothing method \cite{Nesterov2015}.
We remark that even when $\nb(k)=\nd$, the Bandit method is still randomized because of the
random directions $\ub_k$. 

Whereas a KW method involves $\nb(\nx+1)$ component function evaluations in a single iteration, 
the Bandit method entails only $2\nb$ component function evaluations in a single iteration. 
In common with a KW method, however, the Bandit method requires a selection of
the finite-difference parameter $h$,
a sequence of step sizes $\{\alpha_k\}$,
and a sequence of batch sizes $\{\nb(k)\}$.

\subsection{Adaptive sampling quasi-Newton method}
\label{sec:aqn}

We now consider adaptive sampling quasi-Newton (AdaQN) methods 
\cite{BollapragadaICML18, RBSWshort19,RBSWlong20}, which iteratively construct a local quadratic surrogate model according to the sampled component functions and select
search directions $\db_k$ as an approximate minimizer of the quadratic surrogate
model.
The quadratic surrogate model is updated at every iteration using the differences between current and previously evaluated forward-difference gradient approximations.  
Whereas the KW and Bandit methods considered here use a prescribed sequence of batch sizes $\{\nb(k)\}$, 
AdaQN adaptively increases the  
batch size. 
Different adaptive rules \cite{RBSWshort19,RBSWlong20} will increase the batch sizes differently; and we consider one such rule, called the \emph{norm test}, in this study. 

AdaQN computes a direction $\db_k$ of the form similar to \cref{eq:kw-fs}:
\begin{equation}
\label{eq:adaqn-fs}
\db_k = \frac{\nd}{\nb(k)} \Hb_k\displaystyle  \sum_{i\in\B_k} 
\gb_i(\xb_k;h),
\end{equation}
where $\gb_i$ is defined in \cref{eq:FD} and $\Hb_k$ is a quasi-Newton matrix defining the quadratic surrogate model, 
updated such that $\Hb_{k+1} \vb_k = \xb_{k+1} - \xb_k$, where 
\begin{equation}
\label{eq:adqn-fs-hk}
\vb_k =
\displaystyle  \sum_{i\in\B_k} \Big(\gb_i(\xb_{k+1};h) - \gb_i(\xb_k;h)\Big).
\end{equation} 
Unlike KW, however, the step size $\alpha_k$ in AdaQN is adaptively determined in each
iteration via a \emph{stochastic backtracking line search}
\cite{BollapragadaICML18, RBSWshort19}, an automatic procedure that ensures
sufficient decrease in a sampled function. This procedure 
requires evaluating the currently sampled component functions 
along the direction $\db_k$, with the associated number of such evaluations,
$l_k$, varying at each iteration;  typically, $l_k$ is less than $5$. Observe that in
AdaQN, $2|B_k|(\nx+1)$ component function evaluations are required to compute
$\db_k$ (and, for free, $\vb_k$) and $|B_k|l_k$ component function evaluations are required to compute $\alpha_k$. 

The primary hyperparameter in AdaQN is the initial batch size. All the other hyperparameters associated with AdaQN are set to their default values as specified in \cite{RBSWlong20}.

\section{Case Study: Optimizing Fayans energy density functional}
\label{sec:fayans}

A key pursuit in the understanding of atomic nuclei is a global (i.e., across the nuclear chart)  description of nuclei. For such a task, the microscopic tool of choice
is nuclear DFT rooted in the
mean-field approach \cite{Bender2003}. An effective interaction in DFT is given by the EDF,
whose parameters are adjusted to experimental data.
Over the past decade, increasingly refined EDFs  have been developed, with increasingly complex and computationally expensive computer models; see, for example, \cite{UNEDF0,UNEDF1,UNEDF2}. Because of the expense of these computer models, their calibration has focused largely on point-/optimization-based estimation. Bayesian approaches have been demonstrated with limited (e.g., 200 in \cite{McDonnellPRL15}) model evaluations, with nontrivial failure (associated with convergence at insufficient fidelity levels) rates of roughly 9\% of the designs present even in relatively narrow (in terms of posterior probability) regions of parameter space; see \cite{HigdonJPG15}. 

We focus on calibration of a Fayans EDF
\cite{Fayans1998}, for which computer models have recently been developed and
demonstrated to correct some systematic effects of other state-of-the-art
functionals \cite{Reinhard2017}. This functional form has recently sparked
significant interest, especially in the context of charge radii measurements \cite{Hammen2018,Miller2019,Gorges2019,deGroote2020}.

\begin{table}[htb]
\caption{\label{tab:observableClasses} Nine classes of physical
observables that constitute all observables included in the study
\cite{Reinhard2017}.}
\begin{indented}
\lineup
\item[]\begin{tabular}{lcc}
\br
Class & Symbol & Number of Observables\cr
\mr
Binding Energy                            & $E_B$                       & 63 \cr
Charge Radius                    & $r_{\textrm{ch}}$          & 52 \cr
Diffraction Radius                        & $R_{\textrm{diffr}}$        & 28 \cr
Surface Thickness                         & $\sigma$                    & 26 \cr
Neutron Single-Level Energy               & $\epsilon_{ls,n}$           & 5  \cr
Proton Single-Level Energy                & $\epsilon_{ls,p}$           & 5  \cr
Differential Radii                           & $\delta\langle r^2 \rangle$ & 3  \cr
Neutron Pairing Gap                       & $\Delta E_n$                & 5  \cr
Proton Pairing Gap                        & $\Delta E_p$                & 11 \cr
\mr
& & $\nd = 198$ \cr
\br
\end{tabular}
\end{indented}
\end{table}

\subsection{Problem specification}

The computer model $m\left(\nub;\xb\right)$  for the currently used Fayans EDF has $\nx=13$ free model
parameters and employs a pool of fit data for spherical nuclei that primarily comprises
bulk properties of the nuclear ground state (energy, radii, surface thickness),
three-point binding energy differences to calibrate pairing strengths, and some
isotopic differences of  root mean square  radii in calcium isotopes.  Specifically, the pool
used for this study is that used to fit the new Fayans EDF
Fy($\Delta r$) reported in \cite{Reinhard2017} but with the even-odd staggering
of binding energies replaced by the even-even data. 
The total dataset
consists of $\nd=198$ observables of different classes (see
Table~\ref{tab:observableClasses}) that are associated with 72 different spherical, ground-state,
even-even nucleus configurations (with these configurations encapsulated by $\nub$).  
The weights ($\sigma_i$) associated with each observable in the pool are related to those in \cite{Reinhard2017} and are detailed in the 
supplemental material \cite{suppl}. The data, weights, and model outputs together define the collection of component functions $F_1(\xb), \ldots, F_{198}(\xb)$ used in \cref{eq:genfun}.

To ensure that our optimizations solve the specified problem, we identified 
transient platform errors,
transient software faults outside of our control, user error, and 
reproducible software faults in the Fayans model evaluation software
as the classes of failures that can occur during an optimization run and that
must be understood and handled sensibly throughout the study.  
We developed scripts to scan over
all optimization results and their associated metadata so that possible failures
could be flagged and manually inspected.  When a transient failure was positively
identified and  was determined to affect the data quality, the
associated optimization run could simply be rerun and the results manually verified as
acceptable.  Since the failures associated with the Fayans model software, which are discussed further in
\cref{sec:FayansSW}, are reproducible, rerunning a failed optimization is not an option.  As
a result, schemes for handling this class of errors were developed and
implemented.  A detailed discussion of this handling is given in \cref{sec:mods}.

\subsection{Fayans model evaluation software}
\label{sec:FayansSW}
The code that is used in our study to evaluate the Fayans 
EDF is derived from a code solving nonrelativistic nuclear 
Hartree-Fock equations for spherically symmetric nuclei 
\cite{Rei91aR}, which is under continuous development.  We have
identified two classes of reproducible software faults within
the Fayans model evaluation software.  The numerical methods used
internally by the code are iterative, and therefore the first class of
failures is the inability of the methods to satisfy 
a specified stopping criterion within a given maximum
iteration budget. 
While a single computation that does not satisfy
this criterion would normally be deemed as a failed result, for this study and informed by the experience and knowledge of
the developer, we implemented a secondary stopping criterion.  This
criterion, which is a relaxation of the primary
criterion, is employed as follows.  If a computation has failed to
achieve the primary stopping criterion within the budget but does
achieve the secondary criterion within the budget, then the result is
flagged as \emph{marginally convergent}.  If, however, a computation
does not satisfy either criterion within the budget, the associated
model evaluation is flagged as \emph{nonconvergent}.

The second class of failures contains those computations that could not
be run to completion because at runtime the code encountered a situation that
was incompatible with its computations.  Such failures, which are referred to as
\emph{runtime failures}, could arise because 
of exceptional conditions that cause internal algorithmic failures or because 
the computation is being
evaluated in a region of the parameter space for which the functional 
is unstable \cite{Hellemans2013,Pastore2015}.
When runtime
failures occur, the Fayans model code reports an error, execution is aborted,
and the associated model evaluation result is flagged as failed.
To avoid such severe failures as much as possible, we have
established empirical, ``reasonable'' bounds for the model parameters,
where reasonable means that we want to avoid 
instabilities as well as unphysical results (e.g., unbound
matter).  For details regarding the region of assumed stability of Fayans EDF  that is characterized by these bounds, see the supplemental material \cite{suppl}.

Knowledge of this region has not been programmed
into the optimization methods, and therefore any optimization
can evaluate the model at points outside the region of stability.  We expect that
some methods, such as the randomized methods, might have a greater propensity
for choosing points outside the region of stability and that the various methods might also differ in their ability to recover from such bad evaluations.  Our means for managing such potential difficulties is detailed next.

\subsection{Modifications to minimize and address possible failures}
\label{sec:mods}

A necessary step in facilitating the automatic training of any simulation-based model is to ensure that error handling is adequately addressed. All of the methods of \cref{sec:algorithms} use some form of the output  
$\left \{F_i(\xb) ; \; i \in B_k \right\}$
at a queried point $\xb$ to inform their internal decision-making.  Consequently, it is necessary to address what occurs if the evaluation $F_i(\xb)$ fails for one or more components $i \in B_k$. 
In this paper, we seek to make minimal changes to the methods stated in
\cref{sec:algorithms} and instead modify the objective function to
account for the variety of situations that can be encountered as discussed in
\cref{sec:FayansSW}.  
Before detailing each of these modifications,
we stress that throughout this article, information about specific points in parameter
space is reported in the original unscaled space used by physicists.  However,
the optimization methods used in this study were implemented to work on a scaled version of the
parameter space; hence, it is understood that the domain of the
objective function is the scaled space. Unless otherwise stated, the points in parameter
space discussed in the remainder of this section should be assumed to be with respect to
the scaled parameter space used for optimization.  For more information regarding the choice of 
scaling, we refer the reader to the supplemental material \cite{suppl}.

\subsubsection{Projection.}
\label{sec:projection}

The tested optimization methods all were intended primarily for unconstrained optimization. We operate in such a setting here and do not provide any method with prior knowledge 
of valid combinations of parameters.
We observed that, depending on the quality of gradient estimators obtained by the randomized 
methods, the directions $\db_k$ could become so large as to generate steps into physically
meaningless or unstable regions of parameter space.  
To help such methods avoid divergence, we alter the objective function to include a projection 
onto an $\ell_1$-ball centered around the
point $\bar{\xb}$.
The unscaled version of this point is
given in the supplemental material \cite{suppl}.
Because of the scaling, it is
appropriate to use an isotropic $\ell_1$-ball 
for defining a reasonable region; that is, we compute the projection  
\begin{equation}
\label{eq:project}
\xb_{\Pb} = \displaystyle\arg\min_{\yb\in\Pb} \|\yb-\xb\|_2, 
\quad \textrm{where } 
\Pb= \left\{\yb\in \R^{\nx}: \|\yb-\bar{\xb}\|_1\leq 2\right\}.
\end{equation} 
Our choice of using the $\ell_1$-norm to define $\Pb$ is motivated by  
our observation that failures are more likely to occur when many parameter components deviate significantly from $\bar{\xb}$.

We then pass the projected point $\xb_{\Pb}$ to the Fayans model simulation for
evaluation. 
We modify the
objective function \cref{eq:genfun} by applying a multiplicative penalty to each residual $F_i(\xb)$ based on the distance between $\xb_{\Pb}$ and $\xb$; that is,
\begin{equation}
\label{eq:projection_modification}
\tilde{F_i}(\xb) = F_i(\xb_{\Pb})\left(1+ \left\|\xb-\xb_{\Pb}\right\|_2^2\right).
\end{equation}
We acknowledge that the replacement of each $F_i$ with $\tilde{F_i}$
can introduce nonsmoothness 
at the boundary of $\Pb$, even when we assume each $F_i$ is smooth in a neighborhood of the boundary of $\Pb$. 

\subsubsection{Observable convergence.}
\label{sec:convergence}
To account for marginally convergent and noncovergent results as well as
occasional runtime failures, and
informed by the belief that convergent computations are more likely to
indicate physically meaningful points in parameter space, we further modified
the observable data $\tilde{F_i}(\xb)$ in \cref{eq:projection_modification} by
computing
\begin{equation*}
\hspace{-55pt}
\label{eq:convergence_modification}
\hat{F_i}(\xb) = 
\left\{ 
\begin{array}{ll}
\tilde{F_i}(\xb) &\mbox{if the computation of }F_i\left(\xb_{\Pb}\right)\mbox{ succeeded} \\
(1 + \sigmam^2)\tilde{F_i}(\xb) & \mbox{if the computation of }F_i\left(\xb_{\Pb}\right)\mbox{ was marginally convergent} \\
(1 + \sigmat^2)\tilde{F_i}(\xb) & \mbox{if the computation of }F_i\left(\xb_{\Pb}\right)\mbox{ was nonconvergent } \\
(1 + \sigmar^2)\tilde{F_i}(\xb) & \mbox{if the computation of }F_i\left(\xb_{\Pb}\right)\mbox{ had a runtime failure}, 
\end{array}
\right.
\end{equation*}
where $\sigmar \ge \sigmat \ge \sigmam \ge 0$ denote penalty parameters. In our study, we
set $\sigmam=2, \sigmat=5$, and $\sigmar=100$. With these considerations, our 
modified objective function, seen by all of the optimization algorithms, is
\begin{equation}
\label{eq:modfun}
\hat{f}(\xb) = \sum_{i=1}^{\nd} \hat{F_i}(\xb)^2.
\end{equation}

\subsubsection{Recovering from failed simulations.}
\label{sec:failures}
In our study, not even the use of the modified objective function
$\hat{f}(\xb)$ can cover every possible failure case. 
When the Fayans simulation returned no output $\hat{f}(\xb)$ whatsoever---a situation that we refer to as a \emph{hard failure}---none of the methods that we tested can continue. 
We thus slightly modified the methods to handle hard failures. 
The deterministic methods (POUNDERS and Nelder-Mead) were modified to terminate gracefully when a hard failure occurred,
returning the point in parameter space corresponding to the best-found objective value in the run up until the hard failure occurred. 
The randomized methods were augmented with a simple backtracking procedure, like the one employed in AdaQN. 
After a direction $\db$ was computed, if the function evaluation at the next suggested point $\xb+\db$ resulted in hard failure, then the direction $\db$ was replaced by $0.1\db$, and we reevaluated at $\xb+\db$. This process was repeated until the evaluation of $\xb+\db$ did not result in hard failure. 
As we will see in the numerical results, the deterministic methods and AdaQN
never suggested a point that resulted in hard failure; but KW and Bandit did encounter hard failures, depending on the selection of hyperparameters.

\section{Numerical Results}
\label{sec:numerical}

We now study the performance of the algorithms from \cref{sec:algorithms} on the function in \cref{eq:modfun}.
We first tune the identified hyperparameters to obtain hyperparameter values to maximize the  performance of each algorithm. 
Since computational budgets may limit one's ability to perform comprehensive  hyperparameter tuning, the insensitivity to hyperparameter selection 
(as well as the variability overall) may be a key consideration in selecting an optimization algorithm. 
We report on this sensitivity and perform a thorough study of each algorithm using the best hyperparameter values found.

For  this study, the results for all $\nd=198$ component functions were
stored at each evaluated point, even when an optimization method with $\nb<198$ was not provided 
(or charged for) this full set of component function evaluations.
Storing this information allowed us to
reevaluate, during postprocessing, the true function (i.e., with all 198 component functions) 
for every point queried.

The randomized algorithms of \cref{sec:algorithms} require a forward-difference parameter $h>0$. 
In our computations we use $h=5\cdot 10^{-7}$.
This value was obtained by estimating the noise level in each $F_i^2$ following \cite{more2011ecn}. These noise estimates were then used to determine $h$ following the procedure in \cite{more2011edn}. Although variation was seen across different component functions $i$ and different directions in $\R^{\nx}$, the effect of this variation turned out to be mild, and hence we used a fixed difference parameter for all component functions.

\subsection{Tuning of hyperparameters}
\label{sec:tuning}

For our hyperparameter tuning procedure, we randomly selected 5 starting points from the
same $\ell_1$-ball as in \cref{eq:project}. 
We ran each 
method with a budget of $700\nd$ component function evaluations from each starting point. 
Each randomized method was run with three different seeds from each starting point, while deterministic methods were run once from each starting point. 

The three main classes of hyperparameters, and the ways we chose to vary them, are defined below.

\subsubsection{Step-size hyperparameters.}
Every method that we tested, with the exception of AdaQN, requires some kind of (initial) step-size parameter. 
While POUNDERS and Nelder-Mead require a single radius parameter,
the stochastic approximation methods KW and Bandit require a \emph{sequence} of step-size parameters 
$\{\alpha_k\}$, as seen in \cref{eq:gd_iteration}.
For all four methods, we chose to use a common set of step-size hyperparameters based on the set 
$J\equiv \{3,4,5,6,7\}.$

In the case of POUNDERS, the hyperparameter value $\alpha=2^{-j}, j\in J$ sets the initial trust-region radius; 
 and in the case of Nelder-Mead, the hyperparameter value $\alpha=2^{-j}, j\in J$ sets the initial simplex radius. 
For the stochastic approximation methods, we opted to use a schedule of decaying step sizes
$\alpha_k = 2^{-j}/(k+1), j\in J$. 
Employing such a harmonic sequence as the step-size schedule for stochastic approximation methods is in line with standard convergence theory for those methods. 
We remark again that AdaQN employs adaptive step sizes and hence does not require a step-size hyperparameter. 

\subsubsection{Batch-size hyperparameters.}
Each of the stochastic methods
requires the specification of a batch-size parameter. Recall from \cref{sec:sgd} that a batch is drawn uniformly from the $\nd$ component functions and that such draws are independent from one draw to the next.
In each iteration, KW and Bandit methods require a batch size $\nb$ of component function evaluations to compute a gradient estimator; recall \cref{eq:gradsum}. Following standard practice, we chose to hold $\nb$ constant for these methods. 
While AdaQN adaptively increases $\nb$ during the course of an algorithm, it still requires an initial $\nb$. 
For all three  methods, we used a set of 4 common batch sizes
$$\nb\in\{11,33,99,198\}.$$
We interpreted $\nb$ as the constant batch size for KW and Bandit methods and as the initial batch size for AdaQN. 
Observe that all of our tested $\nb$ divide $\nd=198$, which is helpful for comparing the stochastic methods with the full-batch deterministic methods. 
Moreover, when $\nb=\nd$, 
KW and AdaQN are deterministic methods while Bandit is still a randomized method since it employs a random direction $\ub_k$ in each iteration.

\subsection{Performance metrics}
\label{sec:metrics}
We now discuss various measures of effort in order to compare the performance of the methods.
We label by $f_{s,*}$ the minimum function value evaluated over all runs 
instantiated from the $s$th starting point $\xb_{s,0}$, regardless of method, seed, and all relevant hyperparameters. 
We say that a point
$\xb_{s,k}$ is
\emph{$\tau$-optimal for starting point $\xb_{s,0}$} provided 
\begin{equation}
 \label{eq:tau-optimality}
 \displaystyle\frac{f(\xb_{s,k})-f_{s,*}}{f(\xb_{s,0})-f_{s,*}}\leq\tau.
\end{equation}
A point $\xb_{s,k}$ satisfying \cref{eq:tau-optimality} has achieved a fraction $\tau$ of the best-known decrease from starting point $\xb_{s,0}$.

Our primary measure for any run is the best function value, $\min_{k\leq K} f(\xb_{s,k})$, as a function of the number of points, $K$, evaluated during that run. We often report these results in terms of the number of component functions evaluated, that is, $\sum_{k\leq K} \nb(k)$. This also allows us to track the number of component function evaluations needed to achieve $\tau$-optimality, for a specified value of $\tau\in (0,1)$.
Note that for some values of $\tau$, not all runs may achieve $\tau$-optimality; 
when a run fails to do so, we define the number of component function evaluations it required to attain $\tau$-optimality as the budget of component function evaluations it was given.

\subsection{Results of hyperparameter tuning}
\label{sec:empirical}
\begin{figure}[htb]
\centering
\includegraphics[width=0.92\linewidth]{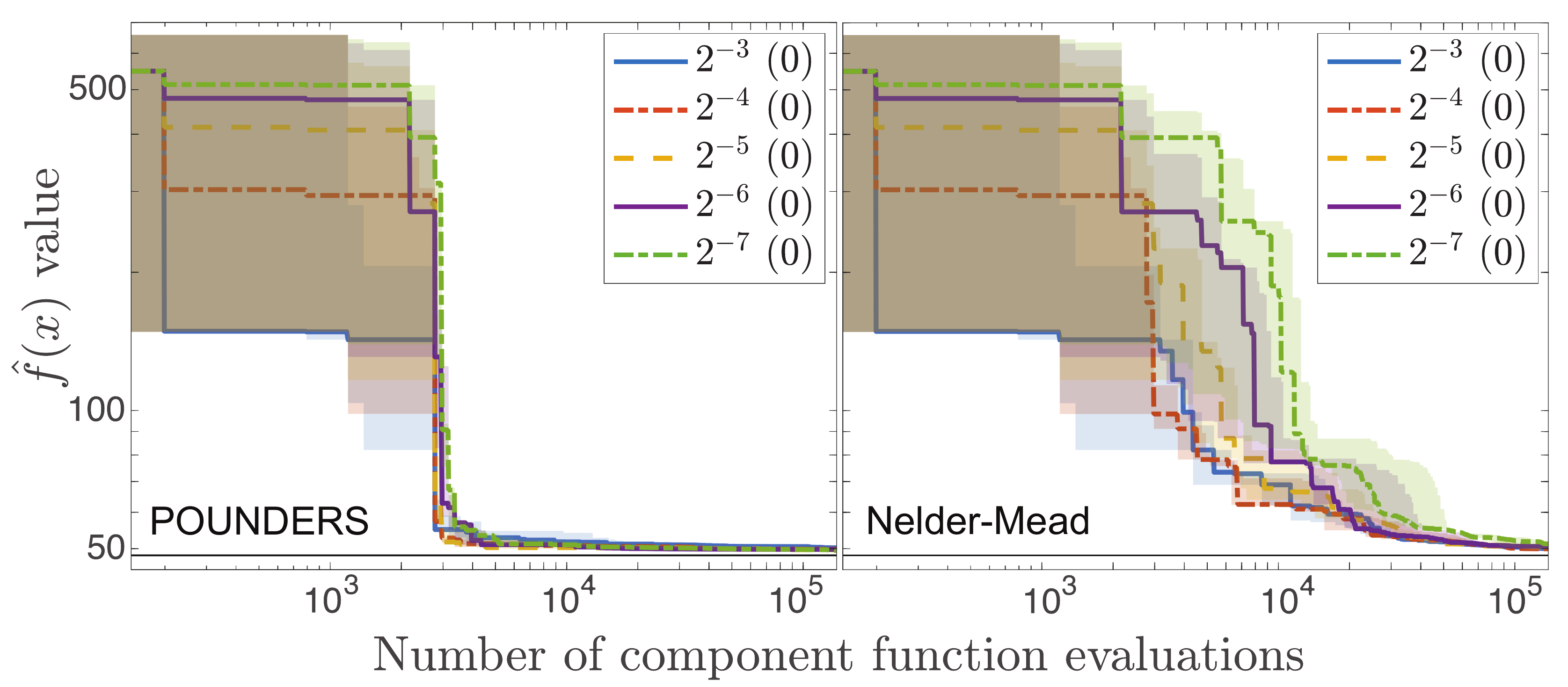}
\caption{Tuning the step-size parameter for POUNDERS (left) and Nelder-Mead (right). 
Throughout such plots in this paper, the 
vertical axis shows the value of $\hat{f}$, which is defined in \cref{eq:modfun}, that is best among those seen within the specified number (horizontal axis) of evaluations. 
Solid lines denote median performance over all starting points (and stochastic replications), while the translucent bands denote the $25$th and $75$th percentiles of performance.
The number in parentheses in the legend denotes the \emph{average} number of hard failures produced by the Fayans-model simulation during the run of the algorithm.
The solid black horizontal line denotes the value of $\hat{f}(\xb_1)$ in Table~\ref{tab:BestResults}. 
\label{fig:pounders_nmead}}
\end{figure}

\begin{figure}[htb]
\centering
\includegraphics[width=.55\linewidth]{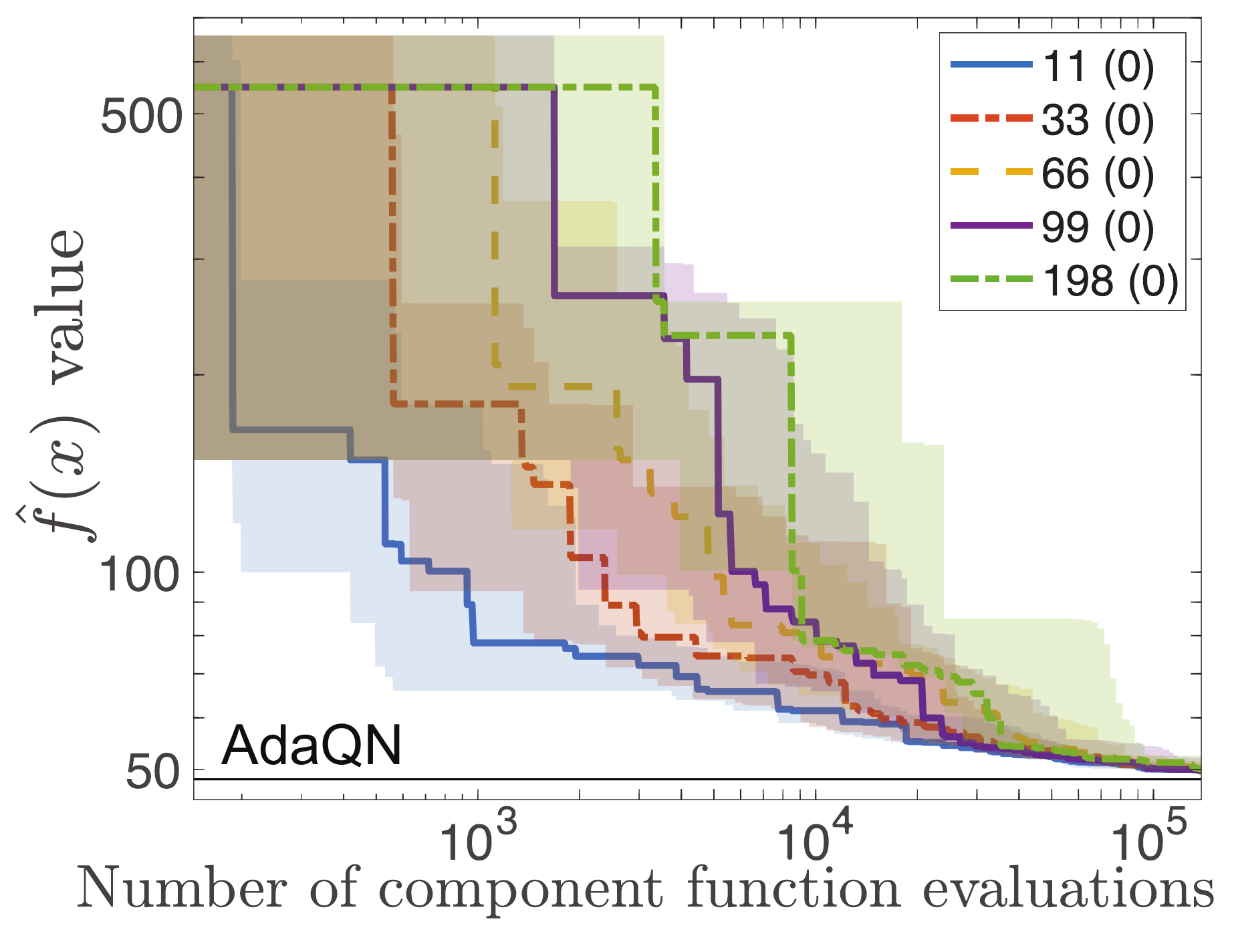}
\caption{Tuning the batch-size parameter for AdaQN. \label{fig:adaqn}} 
\end{figure}

We now show the results of hyperparameter tuning to search for ``best" step sizes and/or batch sizes, where appropriate. 
In \Cref{fig:pounders_nmead} we look at summary hyperparameter tuning results for POUNDERS and Nelder-Mead,
and in \Cref{fig:adaqn} we look at summary results for AdaQN. 
For AdaQN, we 
chose  to tune only initial batch sizes.

Based on the results for AdaQN in \Cref{fig:adaqn}, 
we chose to initialize $\nb=11$, which finds the same quality of median solutions in terms of $\hat{f}$ values as do other batch sizes toward the end of its budget
but identifies better solutions earlier on (in terms of the $25$th, $50$th, and $75$th percentiles). 

\begin{figure}[htb]
\centering
\includegraphics[width=.7\linewidth]{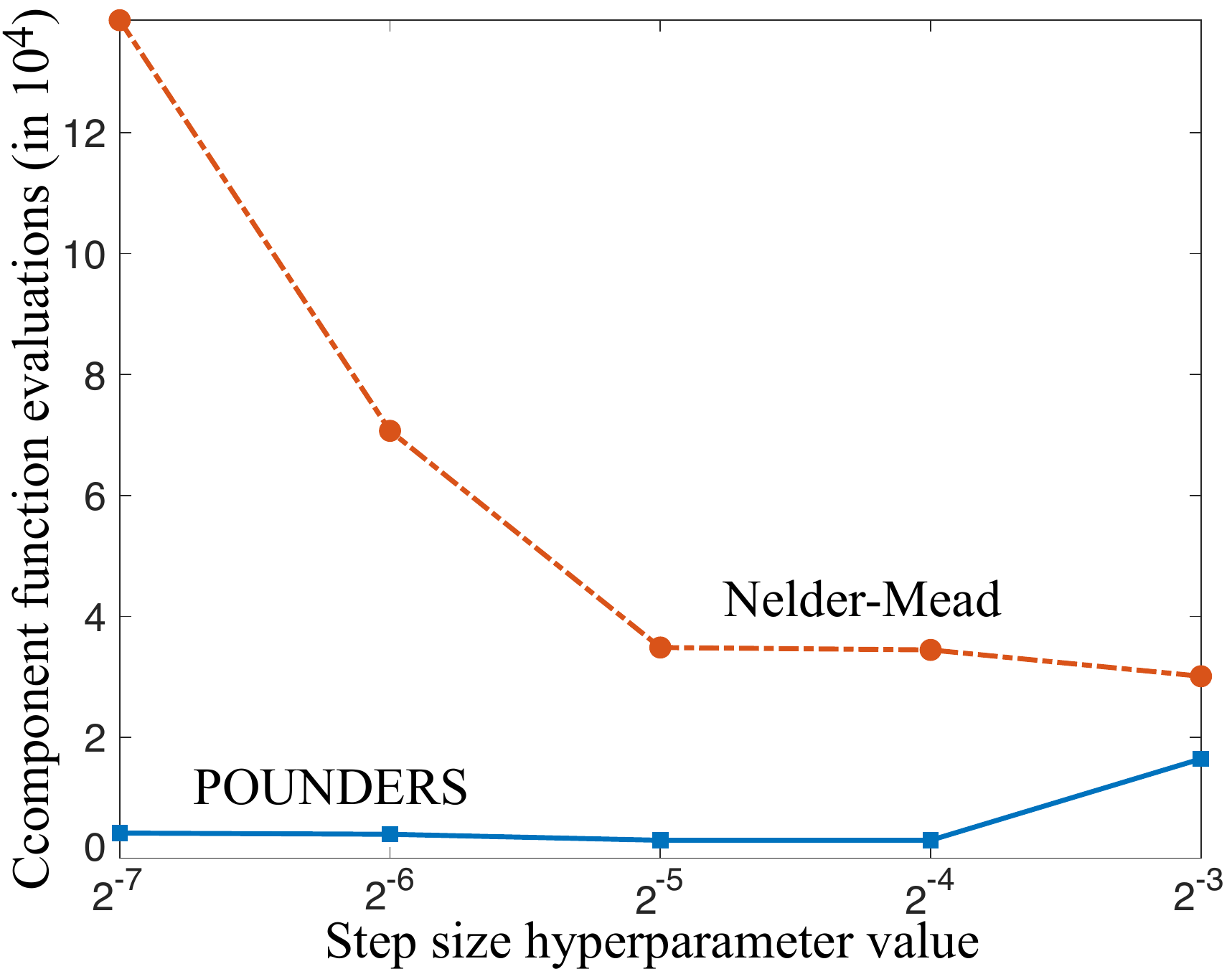} 
\caption{Median number of component function evaluations needed by POUNDERS and Nelder-Mead to attain $\tau$-optimality, where $\tau=0.01$. 
\label{fig:pounders_nmead2}}
\end{figure}

We see in \Cref{fig:pounders_nmead}  that the performance of POUNDERS is extremely robust to the selections of initial trust-region radius. 
For Nelder-Mead, we observe  that its performance is not as independent of the simplex radius as POUNDERS is independent of its initial trust-region radius. This is summarized in \Cref{fig:pounders_nmead2}, which follows the example set in \cite{asi2019importance} and shows
the \emph{median} amount of component function evaluations required by a method to attain $0.01$-optimality. 

Because the POUNDERS performance was so similar for all initial trust-region radius values, we 
selected $\alpha=2^{-4}$. 
For Nelder-Mead, because the best final median function value occurred for $\alpha=2^{-4}$, and because the median performance of $\alpha=2^{-4}$ was nearly as fast as the median performance of $\alpha=2^{-3}$ in finding $0.01$-optimal solutions, we selected $\alpha=2^{-4}$. 

\begin{figure}[htb]
\centering
\includegraphics[width=0.9\linewidth]{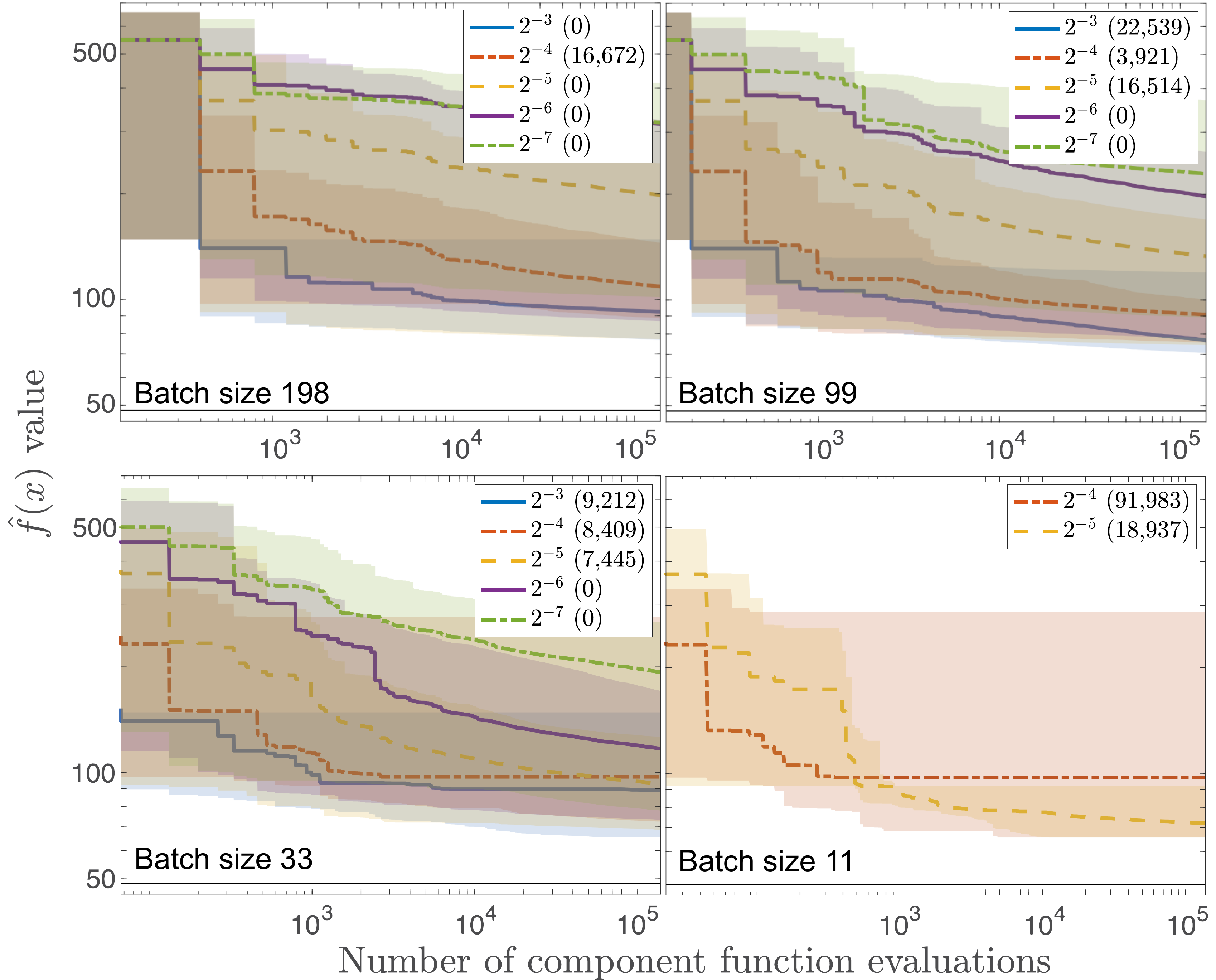}

\caption{Hyperparameter tuning results for Bandit; from left to right, and then top to bottom, are batch sizes 198, 99, 33, and 11.
\label{fig:bandit-batchsizes}}
\end{figure}

\begin{figure}[htb]
\centering
\includegraphics[width=0.9\linewidth]{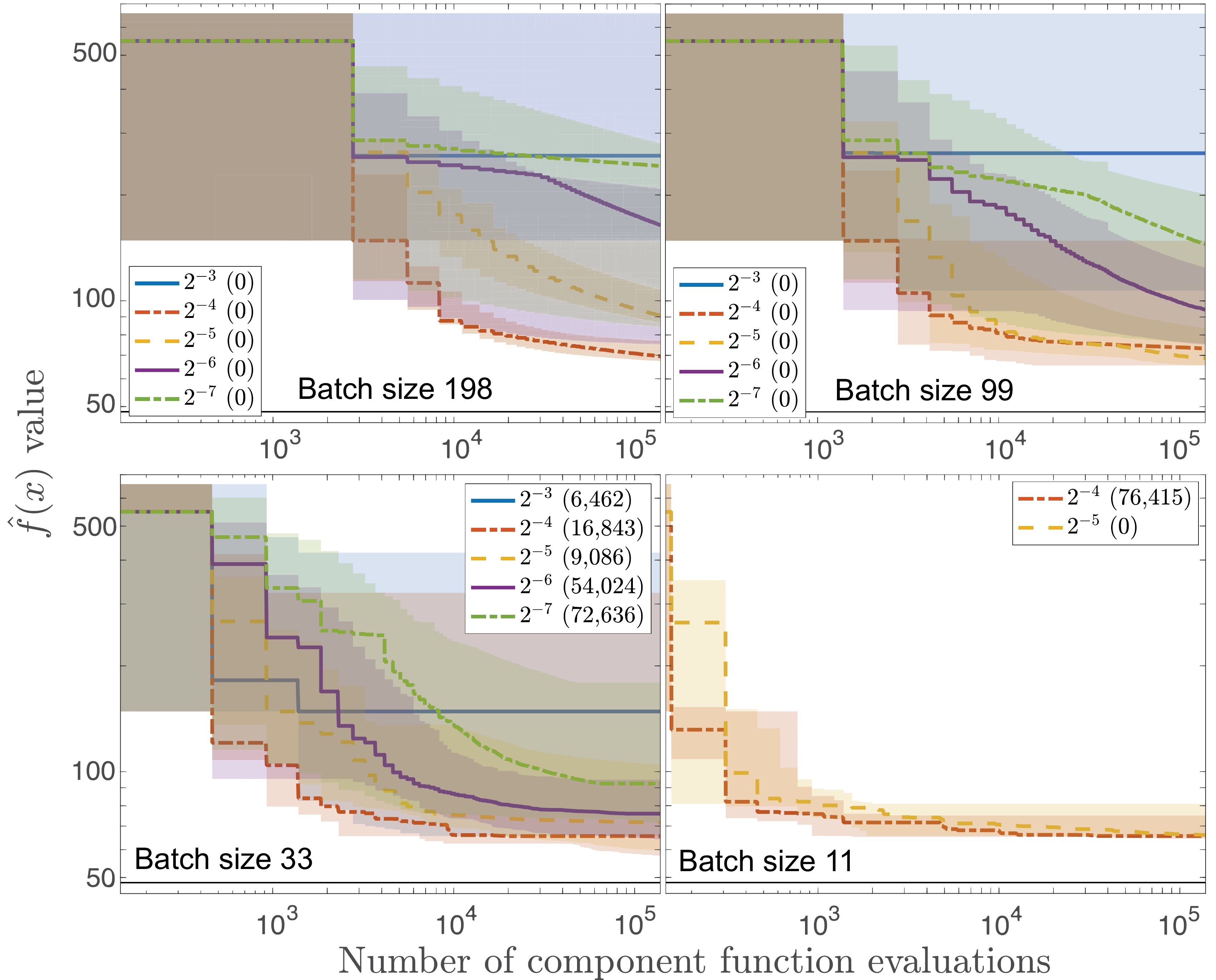}
\caption{Hyperparameter tuning results for KW; from left to right, and then top to bottom, are batch sizes 198, 99, 33, and 11.
\label{fig:kw-batchsizes}}
\end{figure}

For the remaining stochastic approximation methods, we first fix batch-size parameters and compare the median performance of step-size parameters. These results are shown, respectively, in \Cref{fig:bandit-batchsizes} and \Cref{fig:kw-batchsizes}.
We remark that in the case of $\nb=11$, we did not test all the step-size parameters because of the extreme computational cost of running
these two methods with $\nb=11$ in our computational setup. 
Instead, for $\nb=11$ we tested the two step sizes that 
resulted in the fewest average hard failures from running only one seed per starting point.  
In \Cref{fig:kw-batchsizes}, we see that for KW, 
this selection of step sizes matches the selection of step sizes that performed best for $\nb=33$.

In \Cref{fig:bandit-batchsizes}, we observe that for each fixed batch size, there is a step size that provides a clear best median performance. Unlike in the comparisons made for POUNDERS,  Nelder-Mead, and AdaQN, however,
the difference in the percentile performance between the Bandit methods and the average number of hard failures encountered across different runs should be taken into account. 
With these three considerations in mind, for $\nb=198$, we selected $\alpha=2^{-3}$. 
For $\nb=99$, balancing the significantly lower number of hard failures encountered by $\alpha=2^{-4}$ compared with $\alpha=2^{-3}$, as well as the better $75$th percentile performance of $\alpha=2^{-4}$, we selected $\alpha=2^{-4}$. 
For $\nb=33$, because of the similar median final performance of $\alpha=2^{-5}$ and $\alpha=2^{-3}$, coupled with the better $75$th percentile performance of $\alpha=2^{-5}$ and lower number of hard failures encountered by $\alpha=2^{-5}$, we selected $\alpha=2^{-5}$.
For $\nb=11$, the choice of $\alpha=2^{-5}$ was clear. 

In \Cref{fig:kw-batchsizes}, the choice for $\nb=198$ and $\nb=99$ was fairly easy to make, at $\alpha=2^{-4}$ and $\alpha=2^{-5}$, respectively. The choice for $\nb=33$ was less clear; the median performance of $\alpha=2^{-5}$ was not much worse than the median performance of $\alpha=2^{-4}$; however, because the $75$th percentile performance of $\alpha=2^{-5}$ was better than that of $\alpha=2^{-4}$, and because $\alpha=2^{-5}$ had fewer average hard failures, we selected $\alpha=2^{-5}$. 
For $\nb=11$, we selected $\alpha=2^{-5}$ because of its failure-free performance. 

\begin{figure}[htb]
\centering
\includegraphics[width=0.9\linewidth]{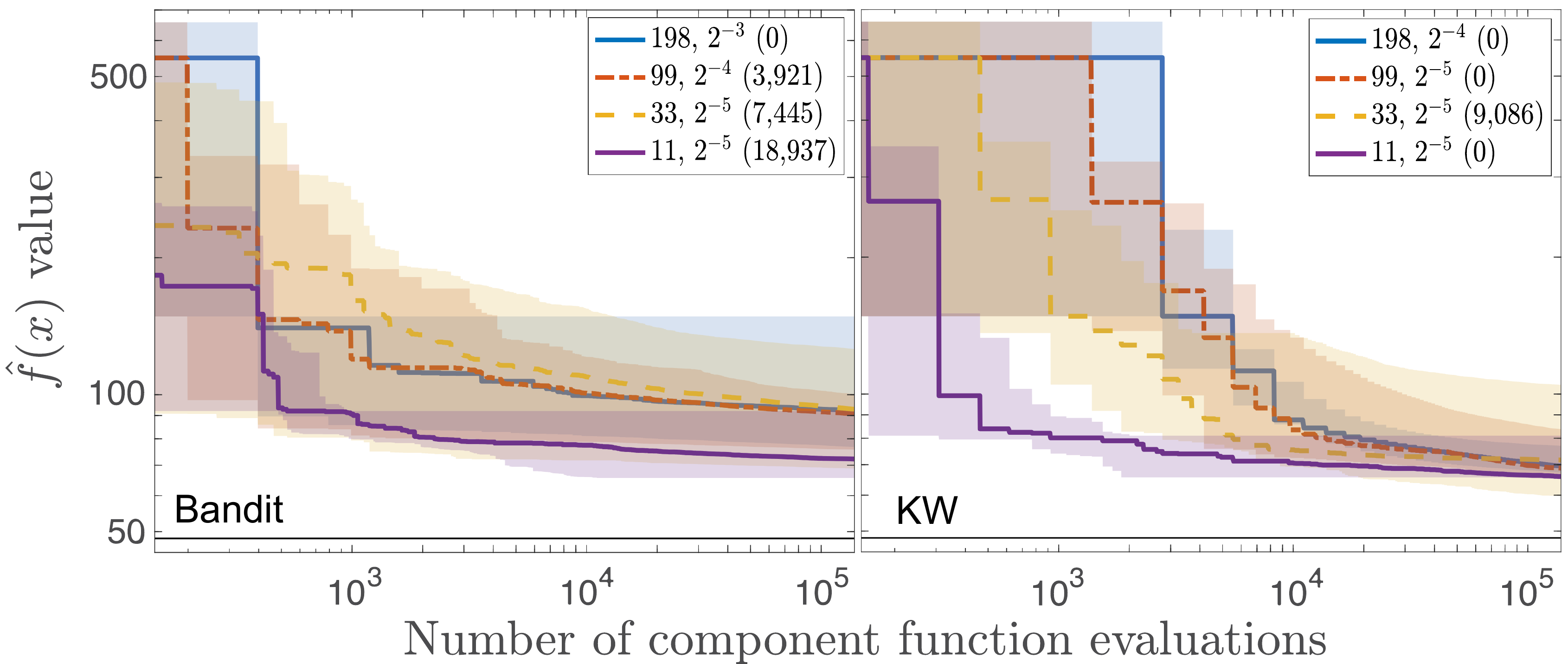}
\caption{Best step sizes of each batch size for Bandit (left) and KW (right). \label{fig:best-stepsizes-compared}}
\end{figure}

Having downselected the step-size parameters per batch size, we now compare the methods across different batch sizes in \Cref{fig:best-stepsizes-compared}.
We see that for KW, using the smallest tested batch size $\nb=11$ with a step size of $\alpha=2^{-5}$ is the best setting of hyperparameters. 
The situation was less clear for Bandit methods. 
While the median performance within the budget of component function evaluations was best with batch size 11, this parameter combination exhibited 
many hard failures; 
as a tradeoff between a smaller number of hard failures and 
a reasonable median performance, we selected step size $\alpha=2^{-4}$ and $\nb=99$.  

\subsection{Comparing tuned methods on additional starting points}
\begin{figure}
\centering
\includegraphics[width=0.6\linewidth]{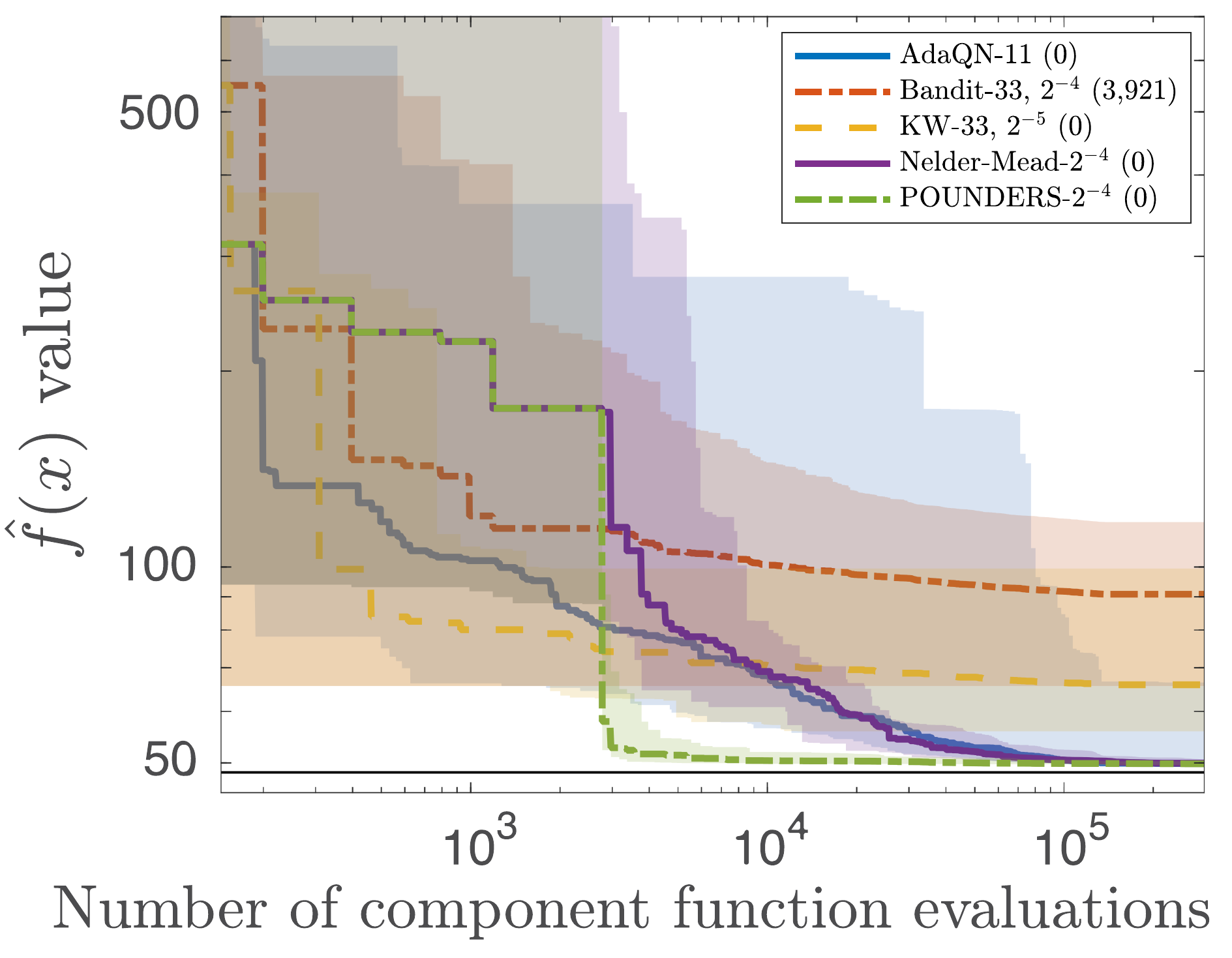}
\caption{Comparing the best variants of all methods. 
 \label{fig:compare-all}}
\end{figure}

Having performed the hyperparameter tuning in the preceding section to select appropriate hyperparameters for each of the five methods, we then ran the selected variant of each method on a larger set of problems.
In particular, we randomly generated twenty starting points (instead of five) from the
same $\ell_1$-ball  as in \cref{eq:project} and again
ran three seeds for each starting point for each of the randomized methods. 
The budget for each method was extended to $1500\nd(\nx+1)$ component function evaluations, more than double the budget provided in the hyperparameter tuning runs.  These results are presented in \Cref{fig:compare-all}.

The results as seen in \Cref{fig:compare-all} are remarkable.
Even in the full run, POUNDERS continues to exhibit an interesting phenomenon where after a fixed number of full-batch component function evaluations, the objective function found suddenly drops and exhibits very low variation in the $25$th to $75$th percentile band in decreasing to a final solution. This sort of robustness and consistency in performance is certainly desirable.

In terms of final median performance, AdaQN and Nelder-Mead find similar quality solutions as POUNDERS. 
One could argue that the performance of Nelder-Mead, in terms of overall median and other percentile trajectories, is 
dominated by the performance of POUNDERS. 
The performance of AdaQN is interesting in that it is not strictly dominated by the performance of POUNDERS. In fact, if one were  interested only in generating reasonable solutions (say $\tau-$optimal solutions, where $\tau\approx 0.25$) in as few component function evaluations as possible, AdaQN is a better choice than POUNDERS. 
If one were simultaneously interested in the robustness of the final solution, then AdaQN remains a strong choice. This is in contrast with KW, which also achieves gains fairly quickly but does not have the same final robustness as exhibited by AdaQN. For all but a few $\tau$ values, the Bandit method is bested by KW, which may be  attributed partly to the nontrivial failure rate experienced by Bandit.

Our comparisons thus far have measured computational expense in terms of the number of component function evaluations. Such metrics are fully justified in computing environments where a single component function can be evaluated at a time.
For sufficiently large parallel computing environments, resources are available to perform full batch (here $\nb=\nd=198$) evaluations simultaneously. We now examine the case between the extremes of 
a single component function evaluated at a time and all $\nd$ component functions evaluated at a time.
Methods capable of subsampling (i.e., using batch sizes $\nb<\nd$) are potentially promising in such intermediate regimes.

Our \emph{resource utilization plots} illustrate these considerations. 
By resource size we denote the number of component function evaluations 
that can be simultaneously computed.
Given a resource size, we refer to the number of \emph{rounds} as the iterations of such whole resources to evaluate the component function evaluations needed to 
achieve a performance metric (e.g., $\tau$-optimality as in \cref{eq:tau-optimality}). 
For example, if the resource size is 11, then a method with $\nb=198$ will use at least 198/11=18 
rounds to make an optimization step, while a method with $\nb=11$ will potentially use only one such round.

\begin{figure}
\centering
\includegraphics[width=0.9\linewidth]{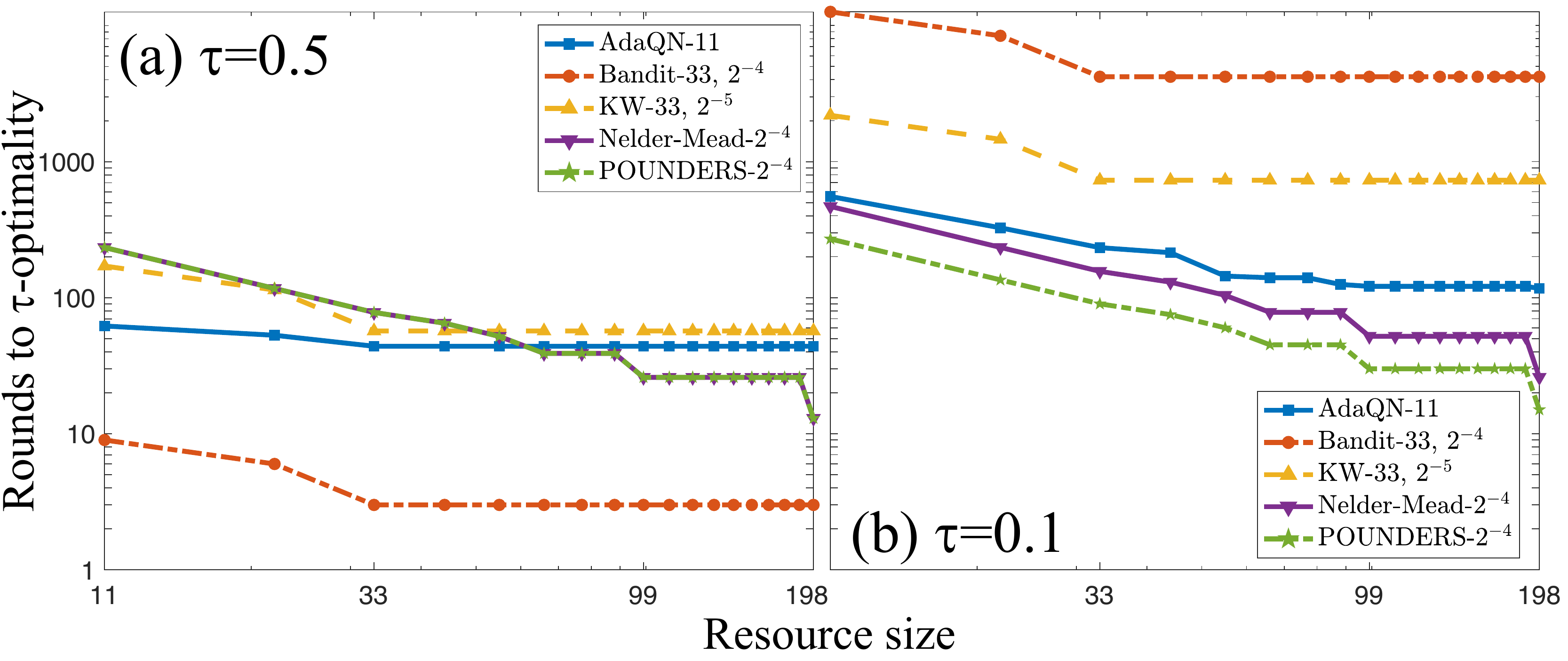}
\caption{Resource utilization plots with respect to $\tau$-optimality (a: $\tau=0.5$; b: $\tau=0.1$) for the tuned methods shown in \Cref{fig:compare-all}. The construction of these figures assumes that component functions can only be evaluated in parallel at a single point in parameter space at a time.  In addition, some choices of resource size result in poor performance when the size is incompatible with $\nb$.  For example, evaluating 33 component functions with a resource size of 22 is charged two rounds (44 possible evaluations).  
Similarly, a method with nonadaptive batch size $\nb$ will not be able to benefit, absent failures, from resource sizes larger than $\nb$.  
 \label{fig:rups}}
\end{figure}

\begin{table}
\caption{\label{tab:initialevals}  Minimum number of initial component function evaluations to evaluate first (noninitialization) optimization step.}
\centering
\begin{tabular}[t]{lr}
\br
  Method & $F_i$ evaluations\\
\mr
AdaQN        & $(\nx+2)\nb(0)$  \\
Bandit       & $3 \nb $ \\
KW           & $(\nx+2)\nb$ \\
Nelder-Mead  & $(\nx+2)\nd$ \\
POUNDERS     & $(\nx+2)\nd$ \\
\br
\end{tabular}
\end{table}

We highlight computational environments where POUNDERS is not the most obvious choice by showing select values of $\tau$ in the resource utilization plots in \Cref{fig:rups}.  
For low demands on solution quality 
($\tau=0.5$), Bandit methods are exceptionally good, identifying $0.5$-optimal solutions remarkably quickly. This is because Bandit requires very few evaluations to get started; \Cref{tab:initialevals} shows that Bandit (with $\nb=33$) will have evaluated its first step (to $\xb_1$) after 99 component function evaluations. The left plot in \Cref{fig:rups} shows that on over half the runs, this number (3 rounds at a resource level of 33) is sufficient for satisfying this coarse accuracy level. Nelder-Mead and POUNDERS show a similar behavior after their first step is performed, but this step requires 15 rounds at a resource level of 198 (90 rounds at a resource level of 33). AdaQN's smaller batch size allows it to outperform POUNDERS and Nelder-Mead at lower resource levels, but is insufficient for catching Bandit at the coarse accuracy $\tau=0.5$.

When we tighten the accuracy demands to $\tau=0.1$, we see that the deterministic methods (i.e., those with a full batch $\nb=198$) are again best, even for resource sizes as small as 11. This plot also shows that AdaQN's adaptive batch size allows it to remain competitive even at this tighter accuracy for resource sizes up to 99.

\section{Discussion}
\label{sec:discussion}

Our results show that the deterministic methods tested were insensitive to starting point in terms of finding a good objective function value with a limited number of $\hat{f}$ evaluations. Furthermore, these methods were generally insensitive to hyperparameters, did not evaluate points that resulted in hard failures, and are attractive even if the expense of evaluating the Fayans model allowed for computing only a fraction (e.g., 11/198=1/18) of the component functions at a time. For problems where even smaller fractions are possible or when less accurate solutions are desired with even smaller computational budgets than those tested here, AdaQN appears especially promising. 
We expect that such methods that can use smaller batch sampling will become more attractive as the number of fit data significantly increases (as in the case of traditional supervised learning applications).

As part of understanding the quality of results achieved in this study, we identified the best run, 
in terms of lowest $\hat{f}$ value, for each of the 20 starting points.
Eleven of these 20 best results were found with POUNDERS, which also found the overall best result; seven by AdaQN; and two by Nelder-Mead.  
All 20 points of these best results are contained in $\Pb$ and resulted in fully converged Fayans model evaluations.
\revised{The parameter values of two of these best points are presented in unscaled form in Table~\ref{tab:BestResults}.}
{The parameter values of two of these best points are presented in
unscaled form in Table~\ref{tab:BestResults}. To give an impression of typical
parameters from previous fits, we also show the parameters for
Fy($\Delta r$) \cite{Reinhard2017} and Fy($\Delta r$,HFB) \cite{Miller2019,Reinhard2020}.}

\begin{table}
\caption{\label{tab:BestResults}  Unscaled parameter values for two of the 20 best optimization results in the study.
\revised{and their associated objective function value.}
{The parameters are given up
to six digits, which suffices to reproduce the output values shown in \Cref{tab:Chi2Comparison}.}
The point $\xb_1$
had the lowest objective function value in the study and is chosen as the representative of
the group of the four best runs; the point $\xb_{5}$ had the best result in
\revised{the second grouping of the remaining 16 best runs. For the definition of Fayans EDF parameters, see \cite{suppl}.  $\rho_\mathrm{eq}$ is in fm$^{-3}$;  $E/A, K, J, L$ are in MeV; other parameters are dimensionless. For comparison, parameter values of Fy($\Delta r$) \cite{Reinhard2017} and Fy($\Delta r$,HFB) \cite{Miller2019,Reinhard2020} EDFs are also given.}
{the second grouping of the remaining 16 best runs. For the definition
of Fayans EDF parameters, see \cite{suppl}.  $\rho_\mathrm{eq}$ is in
fm$^{-3}$;  $E/A, K, J, L$ are in MeV; other parameters are
dimensionless. As a guideline for typical model parameters, the values
for Fy($\Delta r$) \cite{Reinhard2017} and Fy($\Delta r$,HFB)
\cite{Miller2019,Reinhard2020} EDFs are also given.} 
}
\begin{indented}
\lineup
\item[]
\begin{tabular}[t]{lll|ll}
\br
  Parameter & $\xb_1$ & $\xb_{5}$   & \multicolumn{1}{c}{Fy($\Delta r$)}
   & \multicolumn{1}{c}{Fy($\Delta r$,HFB)} \\
\mr
$\rho_\mathrm{eq}$ 	\cmmnt{RHO\_NM}   & \0\00.165755   & \0\00.166182 & \0\00.160 & \0\00.164\\
$E/A$                     \cmmnt{EOVERA}  & \0\-15.8715    & \0\-15.8780 & \0\-16.11 & \0\-15.86 \\
$K$ 	                    \cmmnt{COMPR} & 192.686        & 185.156   & 219 & 210.3\\
$J$ 	                    \cmmnt{ASYMM} & \028.8018      & \028.8467  & \029 & \028.1 \\
$L$ 	                    \cmmnt{DASYM} & \035.6545      & \031.5877  & \030 & \037.5\\
${h_{2-}^\mathrm{v}}$      \cmmnt{H2VM}     & \0\07.08066    & \0\04.71124  & \0\01.2150 & \022.8090 \\
${a_+^\mathrm{s}}$	     \cmmnt{ASP}  & \0\00.594920   & \0\00.620893 & \0\00.6047 & \0\00.56548 \\
${h_{\nabla}^\mathrm{s}}$	    \cmmnt{HGRADP}& \0\00.510148   & \0\00.613192 & \0\00.6656 & \0\00.45795\\
${\kappa}$	          \cmmnt{C0NABJ}  & \0\00.192851   & \0\00.191370 & \0\00.18792 & \0\00.19833 \\
${\kappa'}$	          \cmmnt{C1NABJ}  & \0\00.0383998  & \0\00.0532395 & \0\0\-0.0237 & \0\00.44008 \\
${f_{\mathrm{ex}}^\xi}$	  \cmmnt{FXI}     & \0\0\-3.70050  & \0\0\-3.63760 & \0\0\-4.4720 & \0\0\-4.4556 \\
${h_\nabla^\xi}$	          \cmmnt{HGRADXI} & \0\03.17494    & \0\03.48559 & \0\03.227 & \0\03.113 \\
${h_{+}^\xi}$	          \cmmnt{H1XI}    & \0\03.22592    & \0\03.13267 & \0\04.229 &  \0\04.2440\\[5pt]
\br
\end{tabular}
\end{indented}
\end{table}

\begin{figure}
\centering
\includegraphics[width=.5\linewidth]{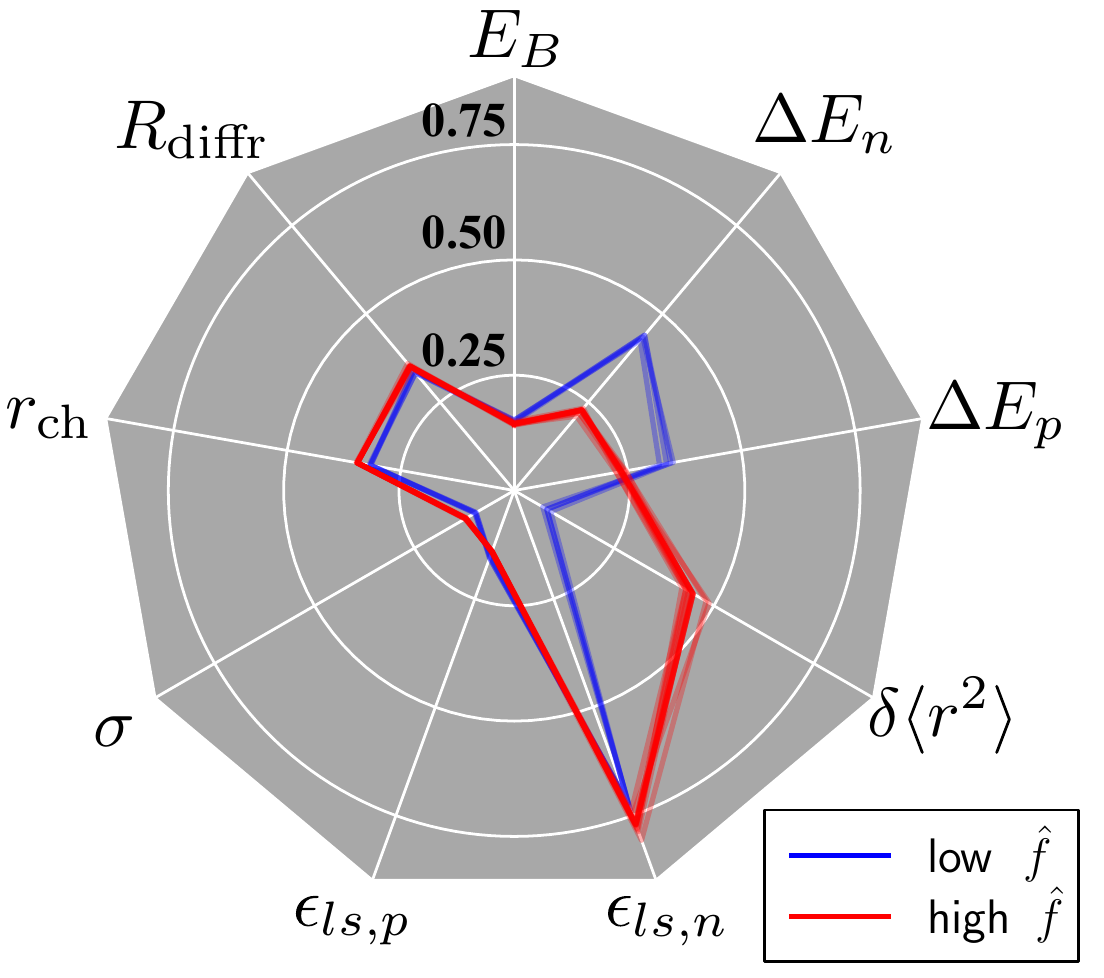}
\caption{Average $\chi^2$ by observable class ($\overline{\chi^2}$) plotted for each of the 20 best
results obtained in the study.  In terms of this quantity, the 20 results clearly can be
partitioned into two different groups.  The results in one such group are
colored blue and also correspond to the four results with the lowest $\hat{f}$
results in the study.\label{fig:MeanChi2PerClass}}
\end{figure}

\Cref{fig:MeanChi2PerClass} shows the outputs of these 20 points by observable class  (see Table~\ref{tab:observableClasses}). For each observable class, by $\chi^2$ we denote the contributions to $\hat{f}(\xb)$ from that observable class (and hence the sum over all observable classes is $\hat{f}(\xb)$). We normalized these $\chi^2$ by the number of observables in the associated class to obtain the average $\chi^2$ of each observable class, $\overline{\chi^2}$. 
\Cref{fig:MeanChi2PerClass} suggests that the results can be partitioned into two groups.  This partitioning is  related not only to $\overline{\chi^2}$ but also to the values of $\hat{f}$.  The four results labeled as Low $\hat{f}$ corresponds to those results with $\hat{f}$ less than 49; the 16 other results, labeled as High $\hat{f}$, have a slightly higher $\hat{f}$.  The results with lowest $\hat{f}$ from each group are denoted by $\xb_1$ and $\xb_5$ in Table~\ref{tab:BestResults}.

The $\chi^2$ and $\overline{\chi^2}$ values are given for the same two points in Table~\ref{tab:Chi2Comparison}.  In general,  the low-$\hat{f}$ cluster appears to fit radius-based observables better than does the high-$\hat{f}$ cluster, but at the expense of the quality of fit to energy-based observables.  In particular, the fit to the isotopic differences of charge radii in Ca isotopes is better, but the fits to the two pairing gaps deteriorate.

\begin{table}
\caption{\label{tab:Chi2Comparison} Breakdown of the $\chi^2$ and average $\chi^2$ ($\overline{\chi^2}$) by 
observable class (see
Table~\ref{tab:observableClasses}) for the two points in parameter space given
in Table~\ref{tab:BestResults}.  In bold are those values with potentially significant differences between the two groups of best results.} 
\begin{indented}
\lineup
\item[]
\begin{tabular}[t]{l||ll|lll}
\br
 & \multicolumn{2}{c|}{$\xb_1$}       &             \multicolumn{2}{c}{$\xb_5$}  \\
Class & $\chi^2$ & $\overline{\chi^2}$ & $\chi^2$ & $\overline{\chi^2}$\\
\mr
$E_B$                       & \09.64 &         0.153  & \09.06 &         0.144  \\
$R_{\rm diffr}$             & \09.49 &         0.339  & \09.81 &         0.351  \\
$r_{\rm ch}$               &  16.41 &         0.316  &  17.95 &         0.345  \\
$\sigma$                    & \02.48 &         0.095  & \03.17 &         0.122  \\
$\epsilon_{ls,p}$           & \00.78 &         0.156  & \00.70 &         0.141  \\
$\epsilon_{ls,n}$           & \03.64 &         0.728  & \03.90 &         0.780  \\
$\delta\langle r^2 \rangle$ & \00.25 & \textbf{0.082} & \01.31 & \textbf{0.436} \\
$\Delta E_p$                & \03.52 & \textbf{0.320} & \02.66 & \textbf{0.242} \\
$\Delta E_n$                & \02.12 & \textbf{0.425} & \01.12 & \textbf{0.225} \\
\mr
$\hat{f}$	            & 48.33  &                & 49.69  & \\
\br
\end{tabular}
\end{indented}
\end{table}

These results underscore the value of optimization methods being able to train physics models with few model evaluations. 
Such efficiency allows one to perform several different optimizations (e.g., from different starting points, with different fit data) and thereby identify potentially different local minimizers. 
The subsequent study of distinct local minima could be useful; the ability of a solution to model the desired physics often matters more than the final objective function value.

\section{Perspectives}

In this study, we addressed the calibration of the nuclear physics model Fayans EDF using the $\chi^2$-minimization, which can be viewed as a supervised machine learning problem.  
The  model is somewhat computationally expensive and the  
derivative information with respect to the model parameters is not available. To this end, we investigated the strengths and limitations of five algorithmic  families  of  iterative  methods  for  local,  unconstrained derivative-free optimization.  We considered two deterministic and three randomized methods. We analyzed hyperparameter tuning considerations and variability associated with the methods, and illustrated considerations for tuning in different computational settings. In total, nearly a half million CPU core hours were expended for this study, an indication of the infeasibility for doing thorough hyperparameter tuning and comparison for many nuclear physics model training problems.  

For the model considered, we conclude that  the  performance  of  POUNDERS, within a  budget  of  function  evaluations,  is  extremely  robust. The Fayans EDF optimization results obtained in this work are generally consistent with those of
Fy($\Delta r$) \cite{Reinhard2017} and Fy($\Delta r$,HFB) \cite{Miller2019,Reinhard2020} models, see  Table~\ref{tab:BestResults}. In particular, the set $\xb_1$, which performs very well on the $\delta\langle r^2 \rangle$ class appears to be fairly close to Fy($\Delta r$,HFB). The extension of the Fayans model to isovector pairing, suggested in  \cite{Reinhard2017}, will be carried out in the following work, which will also contain  detailed discussion of resulting  quantified nuclear properties.

\section*{Acknowledgments}
The work at Argonne was 
supported by the U.S.\ Department of
Energy, Office of Science, Office of Advanced Scientific Computing
Research, applied mathematics and SciDAC programs under Contract No.\
DE-AC02-06CH11357 and by the NUCLEI SciDAC-4 collaboration.  This work was also supported by the U.S.\ Department of Energy, Office of Science, Office of Nuclear Physics under award numbers DE-SC0013365 (Michigan State University) and DE-SC0018083 (NUCLEI SciDAC-4 collaboration).
We gratefully acknowledge the computing resources provided on Bebop, a 
high-performance computing cluster operated by the Laboratory Computing Resource 
Center at Argonne National Laboratory.

\section*{References}
\bibliography{add_abbrev}
\bibliographystyle{iopart-num}



\section*{Supplementary material}

## 1.1. Basic building blocks: local densities and currents

The Fayans EDF, as the Skyrme EDF, is formulated in terms of local densities and currents:

| Symbol | | Expression | Name |
|---|---|---|---|
| $\rho_q$ | $=$ | $\displaystyle\sum_{\alpha\in q} f_\alpha v_\alpha^2 \lvert\varphi_\alpha\rvert^2$ | density |
| $\boldsymbol{s}_q$ | $=$ | $\displaystyle\sum_{\alpha\in q} f_\alpha v_\alpha^2 \varphi_\alpha^+ \hat{\boldsymbol{\sigma}} \varphi_\alpha$ | spin density |
| $\boldsymbol{j}_q$ | $=$ | $\displaystyle\Im m\left\{\sum_{\alpha\in q} f_\alpha v_\alpha^2 \varphi_\alpha^+ \boldsymbol{\nabla}\varphi_\alpha\right\}$ | current |
| $\mathbb{J}_q$ | $=$ | $\displaystyle-\mathrm{i}\sum_{\alpha\in q} f_\alpha v_\alpha^2 \varphi_\alpha^+ \boldsymbol{\nabla}\otimes\hat{\boldsymbol{\sigma}}\varphi_\alpha$ | spin-orbit density |
| $\tau_q$ | $=$ | $\displaystyle\sum_{\alpha\in q} f_\alpha v_\alpha^2 \lvert\boldsymbol{\nabla}\varphi_\alpha\rvert^2$ | kinetic-energy density |
| $\boldsymbol{\tau}_q$ | $=$ | $\displaystyle-\mathrm{i}\sum_{\alpha\in q} f_\alpha v_\alpha^2 \boldsymbol{\nabla}\varphi_\alpha^+ \cdot \boldsymbol{\nabla}\hat{\boldsymbol{\sigma}}\varphi_\alpha$ | kinetic spin-density |
| $\xi_q$ | $=$ | $\displaystyle\sum_{\alpha\in q} f_\alpha u_\alpha v_\alpha \lvert\varphi_\alpha\rvert^2$ | pairing density |

$$(1)$$

In equation (1), $q$ labels the nucleon species with $q = p$ for protons and $q = n$ for neutrons. The $v_\alpha$ and $u_\alpha$ are the standard BCS (or canonical HFB) amplitudes. The $f_\alpha$ is a phase-space weight that provides a smooth cutoff of the space of single-particle states included in pairing. All of the above expressions are local quantities that depend on the position vector $\boldsymbol{r}$ and refer to the local wave function components $\varphi_\alpha = \varphi_\alpha(\boldsymbol{r})$. It is advantageous to handle the densities in terms of isospin. Thus we will consider the recoupled forms, which read for the local density:

$$\rho_+ \equiv \rho = \rho_p + \rho_n, \qquad \rho_- = \rho_p - \rho_n, \tag{2}$$

and similarly for the other densities and currents. The isoscalar density $\rho_+ \equiv \rho$ is equivalent to the total particle density and the difference $\rho_-$ corresponds to the isovector particle density.

The Fayans EDF scales densities with respect to bulk equilibrium density $\rho_{\mathrm{sat}}$. Thus we will employ in the following the density in terms of

$$x_+ = \frac{\rho_+}{\rho_{\mathrm{sat}}}, \qquad x_- = \frac{\rho_-}{\rho_{\mathrm{sat}}}. \tag{3}$$

## 1.2. The functional

The Fayans EDF is a nonrelativistic energy density functional similar to the widely used Skyrme functional [1], but has more flexibility in density dependence and pairing. We use it here in the form of the original FaNDF0 parameterization [2]. In the following, we restrict ourselves to terms that are actually used in the present study and we mark free model parameters in <span style="color:magenta">magenta</span> and fixed parameters in <span style="color:green">green</span>. The latter serve mostly

as scaling parameters. The functional then reads as follows.

$$E = E_{\text{kin}} + \int d^3r \left( \mathcal{E}_{\text{Fy}}(\rho, \tau, \boldsymbol{j}, \boldsymbol{J}) + \mathcal{E}_{\text{C,ex}}(\rho_p) + \mathcal{E}_{\text{pair}}(\chi, \rho) \right)$$
$$+ E_C(\rho_p) - E_{\text{cm}} \tag{4a}$$

$$E_{\text{kin}} = \int d^3r \left( \frac{\hbar^2}{2m_p} \tau_p + \frac{\hbar^2}{2m_n} \tau_n \right) \tag{4b}$$

$$\mathcal{E}_{\text{Fy}} = \mathcal{E}_{\text{Fy}}^{\text{v}}(\rho) + \mathcal{E}_{\text{Fy}}^{\text{s}}(\rho) + \mathcal{E}_{\text{Fy}}^{(\text{kin})}(\rho, \tau, \boldsymbol{j}) + \mathcal{E}_{\text{Fy}}^{(\text{ls})}(\rho, \boldsymbol{J}) \tag{4c}$$

$$\mathcal{E}_{\text{Fy}}^{\text{v}} = \tfrac{1}{3} \epsilon_F^{(0)} \rho_{\text{sat}} \left[ a_+^{\text{v}} \frac{1 - h_{1+}^{\text{v}} x_+^{\sigma}}{1 + h_{2+}^{\text{v}} x_+^{\sigma}} x_+^2 + a_-^{\text{v}} \frac{1 - h_{1-}^{\text{v}} x_+}{1 + h_{2-}^{\text{v}} x_+} x_-^2 \right] \tag{4d}$$

$$\mathcal{E}_{\text{Fy}}^{\text{s}} = \tfrac{1}{3} \epsilon_F^{(0)} \rho_{\text{sat}} \frac{a_+^{\text{s}} r_0^2 (\boldsymbol{\nabla} x_+)^2}{1 + h_+^{\text{s}} x_+^{\sigma} + h_{\nabla}^{\text{s}} r_0^2 (\boldsymbol{\nabla} x_+)^2} \tag{4e}$$

$$\mathcal{E}_{\text{Fy}}^{\text{ls}} = \frac{4 \epsilon_F^{(0)} r_0^2}{3 \rho_{\text{sat}}} \left( \kappa \rho \boldsymbol{\nabla} \cdot \boldsymbol{J} + \kappa' \rho_- \boldsymbol{\nabla} \cdot \boldsymbol{J}_- \right) \tag{4f}$$

$$\mathcal{E}_{\text{Fy}}^{\text{pair}} = \frac{4 \epsilon_F^{(0)}}{3 \rho_{\text{sat}}} \sum_{q \in \{p,n\}} \xi_q^2 \left[ f_{\text{ex}}^{\xi} + h_1^{\xi} x_{P+}^{\gamma} + h_{\nabla}^{\xi} r_0^2 (\boldsymbol{\nabla} x_{P+})^2 \right], \ x_{P+} = \frac{\rho_+}{\rho_{0\text{pair}}} \tag{4g}$$

$$E_C = \tfrac{1}{2} e^2 \int d^3r \, d^3r' \rho_C(\boldsymbol{r}) \frac{1}{|\boldsymbol{r} - \boldsymbol{r}'|} \rho_C(\boldsymbol{r}') \tag{4h}$$

$$\mathcal{E}_{\text{C,ex}} = -\tfrac{3}{4} e^2 \left( \frac{3}{\pi} \right)^{1/3} \rho_p^{4/3} \tag{4i}$$

$$\epsilon_F^{(0)} = \left( \frac{9\pi}{8} \right)^{2/3} \frac{\hbar^2}{2 m r_0^2} \tag{4j}$$

$$r_0 = \left( \frac{3}{8\pi \rho_{\text{sat}}} \right)^{1/3} \tag{4k}$$

The center-of-mass correction

$$E_{\text{cm}} = \frac{\langle \hat{P}_{\text{cm}}^2 \rangle}{2mA} \tag{4l}$$

is special in that it is not included in the variational mean-field equations, but is subtracted a posteriori. Note also that the direct term of the Coulomb energy (4h) employs the charge density $\rho_C$, which is the proton and neutron density folded with the intrinsic charge distribution of proton or neutron [3]. The fixed parameters (marked in green) are

$$\frac{\hbar^2}{2m_p} = 20.749811 \, \text{MeV} \, \text{fm}^2$$
$$\frac{\hbar^2}{2m_n} = 20.721249 \, \text{MeV} \, \text{fm}^2$$
$$\rho_{\text{sat}} = 0.16 \, \text{fm}^{-3}$$
$$\sigma = 1/3$$
$$\gamma = 2/3$$
$$e^2 = 1.43996448 \, \text{MeV} \, \text{fm}.$$

## 2. The optimization dataset

The basis of the data used for optimizing the Fayans EDF are binding energies and key properties of the charge form factor [4] such as charge radius, diffraction radius, and surface thickness. The data are shown in Tables 1 and 2. Data points and their adopted errors were chosen in a mutually dependent manner. The errors were chosen such that each class of observables contributed a $\chi^2$ per data point of about one [5, 6]. The nuclear ground-state data points were selected such that the systematic errors from ground-state correlations (beyond DFT) remain smaller than the adopted errors [7]. The additional data points on differential charge radii were given small adopted errors to promote good adjustment of these new data points.

Part of this data basis set are also some single-particle properties, namely a few spin-orbit splitting of single-particle levels. Their uncertainty is given as relative error and taken rather large because single-particle energies are indirectly deduced from experiment.

Like the previous fits of the Fayans EDF [8, 9], the dataset includes three-point staggering of binding energies for calibrating pairing properties, see Table 4. In this case, however, the dataset includes even-even staggering as opposed to even-odd staggering; see [8] for more details.

Finally, a few crucial differences of charge radii (coined differential radii) in Ca are included in the fit data, see Table 4. These were decisive to determine the advanced gradient terms in the Fayans EDF related to the parameters $h_\nabla^s$ and $h_\nabla^\xi$. For a detailed discussion of the physics implications see [8].

**Table 1.** Basic experimental data for the fits together with their adopted error. Part I: along isotopic chains.

| A | Z | $E_B$ | $\Delta E_B$ | $R_{\text{diffr}}$ | $\Delta R_{\text{diffr}}$ | $\sigma$ | $\Delta\sigma$ | $r_{\text{ch}}$ | $\Delta r_{\text{ch}}$ |
|---|---|---|---|---|---|---|---|---|---|
| | | MeV | | fm | | fm | | fm | |
| 16 | 8 | -127.620 | 4 | 2.777 | 0.08 | 0.839 | 0.08 | 2.701 | 0.04 |
| 36 | 20 | -281.360 | 2 | | | | | | |
| 38 | 20 | -313.122 | 2 | | | | | | |
| 40 | 20 | -342.051 | 3 | 3.845 | 0.04 | 0.978 | 0.04 | 3.478 | 0.02 |
| 42 | 20 | -361.895 | 2 | 3.876 | 0.04 | 0.999 | 0.04 | 3.513 | 0.04 |
| 44 | 20 | -380.960 | 2 | 3.912 | 0.04 | 0.975 | 0.04 | 3.523 | 0.04 |
| 46 | 20 | -398.769 | 2 | | | | | 3.502 | 0.02 |
| 48 | 20 | -415.990 | 1 | 3.964 | 0.04 | 0.881 | 0.04 | 3.479 | 0.04 |
| 50 | 20 | -427.491 | 1 | | | | | 3.523 | 0.18 |
| 52 | 20 | -436.571 | 1 | | | | | 3.5531 | 0.18 |
| 58 | 26 | | | | | | | 3.7745 | 0.18 |
| 56 | 28 | -483.990 | 5 | | | | | 3.750 | 0.18 |
| 58 | 28 | -506.500 | 5 | 4.364 | 0.04 | | | 3.776 | 0.10 |
| 60 | 28 | -526.842 | 5 | 4.396 | 0.04 | 0.926 | 0.20 | 3.818 | 0.10 |
| 62 | 28 | -545.258 | 5 | 4.438 | 0.04 | 0.937 | 0.20 | 3.848 | 0.10 |
| 64 | 28 | -561.755 | 5 | 4.486 | 0.04 | 0.916 | 0.08 | 3.868 | 0.10 |
| 68 | 28 | -590.430 | 1 | | | | | | |
| 100 | 50 | -825.800 | 2 | | | | | | |
| 108 | 50 | | | | | | | 4.563 | 0.04 |
| 112 | 50 | | | 5.477 | 0.12 | 0.963 | 0.36 | 4.596 | 0.18 |
| 114 | 50 | | | 5.509 | 0.12 | 0.948 | 0.36 | 4.610 | 0.18 |
| 116 | 50 | | | 5.541 | 0.12 | 0.945 | 0.36 | 4.626 | 0.18 |
| 118 | 50 | | | 5.571 | 0.08 | 0.931 | 0.08 | 4.640 | 0.02 |
| 120 | 50 | | | 5.591 | 0.04 | | | 4.652 | 0.02 |
| 122 | 50 | -1035.530 | 3 | 5.628 | 0.04 | 0.895 | 0.04 | 4.663 | 0.02 |
| 124 | 50 | -1050.000 | 3 | 5.640 | 0.04 | 0.908 | 0.04 | 4.674 | 0.02 |
| 126 | 50 | -1063.890 | 2 | | | | | | |
| 128 | 50 | -1077.350 | 2 | | | | | | |
| 130 | 50 | -1090.400 | 1 | | | | | | |
| 132 | 50 | -1102.900 | 1 | | | | | | |
| 134 | 50 | -1109.080 | 1 | | | | | | |
| 198 | 82 | -1560.020 | 9 | | | | | 5.450 | 0.04 |
| 200 | 82 | -1576.370 | 9 | | | | | 5.459 | 0.02 |
| 202 | 82 | -1592.203 | 9 | | | | | 5.474 | 0.02 |
| 204 | 82 | -1607.521 | 2 | 6.749 | 0.04 | 0.918 | 0.04 | 5.483 | 0.02 |
| 206 | 82 | -1622.340 | 1 | 6.766 | 0.04 | 0.921 | 0.04 | 5.494 | 0.02 |
| 208 | 82 | -1636.446 | 1 | 6.776 | 0.04 | 0.913 | 0.04 | 5.504 | 0.02 |
| 210 | 82 | -1645.567 | 1 | | | | | 5.523 | 0.02 |
| 212 | 82 | -1654.525 | 1 | | | | | 5.542 | 0.02 |
| 214 | 82 | -1663.299 | 1 | | | | | 5.559 | 0.02 |

**Table 2.** Basic experimental data for the fits together with their adopted error. Part II: along isotonic chains.

| A | Z | $E_B$ | $\Delta E_B$ | $R_{\text{diffr}}$ | $\Delta R_{\text{diffr}}$ | $\sigma$ | $\Delta\sigma$ | $r_{\text{ch}}$ | $\Delta r_{\text{ch}}$ |
|---|---|---|---|---|---|---|---|---|---|
| | | MeV | | fm | | fm | | fm | |
| 34 | 14 | -283.429 | 2 | | | | | | |
| 36 | 16 | -308.714 | 2 | 3.577 | 0.16 | 0.994 | 0.16 | 3.299 | 0.02 |
| 38 | 18 | -327.343 | 2 | | | | | 3.404 | 0.02 |
| 42 | 22 | -346.904 | 90 | | | | | | |
| 50 | 22 | -437.780 | 2 | 4.051 | 0.04 | 0.947 | 0.08 | 3.570 | 0.02 |
| 52 | 24 | -456.345 | 90 | 4.173 | 0.04 | 0.924 | 0.16 | 3.642 | 0.04 |
| 54 | 26 | -471.758 | 90 | 4.258 | 0.04 | 0.900 | 0.16 | 3.693 | 0.04 |
| 86 | 36 | -749.235 | 2 | | | | | 4.184 | 0.02 |
| 88 | 38 | -768.467 | 1 | 4.994 | 0.04 | 0.923 | 0.04 | 4.220 | 0.02 |
| 90 | 40 | -783.893 | 1 | 5.040 | 0.04 | 0.957 | 0.04 | 4.269 | 0.02 |
| 92 | 42 | -796.508 | 1 | 5.104 | 0.04 | 0.950 | 0.04 | 4.315 | 0.02 |
| 94 | 44 | -806.849 | 2 | | | | | | |
| 96 | 46 | -815.034 | 2 | | | | | | |
| 98 | 48 | -821.064 | 2 | | | | | | |
| 134 | 52 | -1123.270 | 1 | | | | | | |
| 136 | 54 | -1141.880 | 1 | | | | | 4.791 | 0.02 |
| 138 | 56 | -1158.300 | 1 | 5.868 | 0.08 | 0.900 | 0.08 | 4.834 | 0.02 |
| 140 | 58 | -1172.700 | 1 | | | | | 4.877 | 0.02 |
| 142 | 60 | -1185.150 | 2 | 5.876 | 0.12 | 0.989 | 0.12 | 4.915 | 0.02 |
| 144 | 62 | -1195.740 | 2 | | | | | 4.960 | 0.02 |
| 146 | 64 | -1204.440 | 2 | | | | | 4.984 | 0.02 |
| 148 | 66 | -1210.750 | 2 | | | | | 5.046 | 0.04 |
| 150 | 68 | -1215.330 | 2 | | | | | 5.076 | 0.04 |
| 152 | 70 | -1218.390 | 2 | | | | | | |
| 206 | 80 | -1621.060 | 1 | | | | | 5.485 | 0.02 |
| 210 | 84 | -1645.230 | 1 | | | | | 5.534 | 0.02 |
| 212 | 86 | -1652.510 | 1 | | | | | 5.555 | 0.02 |
| 214 | 88 | -1658.330 | 1 | | | | | 5.571 | 0.02 |
| 216 | 90 | -1662.700 | 1 | | | | | | |
| 218 | 92 | -1665.650 | 1 | | | | | | |

**Table 3.** Experimental data for a few selected spin-orbit splittings.

| A | Z | level | $\varepsilon_{ls,p}$ | $\Delta\varepsilon_{ls,p}$ | level | $\varepsilon_{ls,n}$ | $\Delta\varepsilon_{ls,n}$ |
|---|---|-------|----------------------|----------------------------|-------|----------------------|----------------------------|
|   |   |       | MeV | | | MeV | |
| 16 | 8 | 1p | 6.30 | 60% | 1p | 6.10 | 60% |
| 40 | 20 | 1f | 7.2 | 1800% | 1f | 6.3 | 1800% |
| 48 | 20 | 1f | 4.3 | 1800% | | | |
| 132 | 50 | 2p | 1.35 | 20% | 2d | 1.65 | 20% |
| 208 | 82 | 2d | 1.42 | 20% | 2f | 0.90 | 20% |
| | | | | | 3p | 1.77 | 0.72 |

**Table 4.** The fit data on calcium differential radii is defined as $\delta\langle r^2\rangle^{A,A'} = \delta\langle r_{\mathrm{ch}}^2\rangle^{A'} - \delta\langle r_{\mathrm{ch}}^2\rangle^{A}$; three-point energy differences for even-even ground state nuclei, as $\Delta E_n(Z,N) = \frac{1}{2}(E_B(Z,N+2) - 2E_B(Z,N) + E_B(Z,N-2))$ for neutrons and similarly for protons. The adopted errors are in parentheses with units of fm$^2$ for $\delta\langle r^2\rangle$ and MeV for energy differences.

| Dataset | Fit observables |
|---------|-----------------|
| neutron $\Delta E_n$ | $^{44}$Ca (0.24), $^{118}$Sn (0.36), $^{120}$Sn (0.36), $^{122}$Sn (0.24), $^{124}$Sn (0.24) |
| proton $\Delta E_p$ | $^{36}$S (0.36), $^{88}$Sr (0.36), $^{90}$Zr (0.24), $^{92}$Mo (0.12), $^{94}$Ru (0.24), $^{136}$Xe (0.24), $^{138}$Ba (0.24), $^{140}$Ce (0.24), $^{142}$Nd (0.24), $^{214}$Ra (0.24), $^{216}$Th (0.24) |
| $\Delta r^2$ | $\delta\langle r^2\rangle^{48,40}$ (0.008), $\delta\langle r^2\rangle^{48,44}$ (0.008), $\delta\langle r^2\rangle^{52,48}$ (0.02) |

**Table 5.** Values for the free model parameters considered in our case study. The $\ell_1$-ball described and discussed in the article is centered on the scaled point $\bar{\mathbf{x}}$. In this table, the unscaled version, $\bar{\mathbf{x}}_u$, of this point is given. The lengthscales used to define our scaled space by scaling each model parameter are also given. The lower-bound (LB) and upper-bound (UB) columns specify a rough guess for the limits of stability of the calculations. This means that the region in which the Fayans-model software is believed to be numerically stable is the largest ellipsoid that fits inside the rectangular region specified by the values in these columns. $\rho_{\mathrm{eq}}$ is in $\mathrm{fm}^{-3}$; $E/A, K, J, L$ are in MeV; other parameters are dimensionless.

|  | $\bar{\mathbf{x}}_u$ | Scaling | LB | UB |
|---|---|---|---|---|
| $\rho_{\mathrm{eq}}$ | 0.1642 | 0.004 | 0.146 | 0.167 |
| $E/A$ | $-15.86$ | 0.1 | $-16.21$ | $-15.50$ |
| $K$ | 206.6 | 25 | 137.2 | 234.4 |
| $J$ | 28.3 | 3.2 | 19.5 | 37.0 |
| $L$ | 35.9 | 32 | 2.2 | 69.6 |
| $h_{2-}^{v}$ | 11.34 | 19.01 | 0 | 100 |
| $a_{+}^{s}$ | 0.562 | 0.06 | 0.418 | 0.706 |
| $h_{\nabla}^{s}$ | 0.460 | 0.24 | 0 | 0.516 |
| $\kappa$ | 0.188 | 0.02 | 0.076 | 0.216 |
| $\kappa'$ | 0.045 | 0.17 | $-0.892$ | 0.982 |
| $f_{\mathrm{ex}}^{\xi}$ | $-4.46$ | 1.16 | $-4.62$ | $-4.38$ |
| $h_{+}^{\xi}$ | 4.18 | 1.68 | 3.94 | 4.27 |
| $h_{\nabla}^{\xi}$ | 3.44 | 1.4 | $-0.96$ | 3.66 |

## 3. Parameter scaling and parameter boundaries

The model parameters carry different physical dimensions and different scales of sensitivity. To make them comparable, we need a proper relative scaling. This scaling was achieved by starting from a near-optimal parameter set ($\bar{\mathbf{x}}_u$) and varying each parameter separately to obtain similar orders of magnitude of changes in the output. The length scales used to scale our parameter space are given in Table 5.

The notion of a "free" parameter has to be taken with care. The model parameters cannot be varied freely as there are regions where the parameters produce unphysical output or no solution at all. It is advisable to confine the parameter search to regions where we can expect model stability [10, 11]. This is done in the last two columns of Table 5, which gives for each free parameter a large interval of stability. The intervals were explored by starting from a near-optimal parameter set and varying each parameter separately until results are unphysical or run into unstable solutions. We then confine the parameter variation to a multidimensional ellipsoid touching the given interval bounds for each parameter.



\clearpage
\vspace{3em}

\small

\framebox{\parbox{\linewidth}{
The submitted manuscript has been created by UChicago Argonne, LLC, Operator of 
Argonne National Laboratory (``Argonne''). Argonne, a U.S.\ Department of 
Energy Office of Science laboratory, is operated under Contract No.\ 
DE-AC02-06CH11357. 
The U.S.\ Government retains for itself, and others acting on its behalf, a 
paid-up nonexclusive, irrevocable worldwide license in said article to 
reproduce, prepare derivative works, distribute copies to the public, and 
perform publicly and display publicly, by or on behalf of the Government.  The 
Department of Energy will provide public access to these results of federally 
sponsored research in accordance with the DOE Public Access Plan. 
\url{http://energy.gov/downloads/doe-public-access-plan.}}}

\end{document}